\documentclass[desactivate]{aa} 

\usepackage[varg]{txfonts}
\usepackage{orcidlink}
\usepackage{graphicx}   % Including figure files
\usepackage{hyperref}

\usepackage{natbib}
\bibpunct{(}{)}{;}{a}{}{,}

\newcommand{\response}[1]{{#1}}% responses to referee
\newcommand{\edit}[1]{{#1}}% final edits

\begin{document} 

   \title{Massive star cluster formation }

   \subtitle{I. High star formation efficiency while resolving feedback of individual stars}
   % \subtitle{High Star Formation Efficiency with Resolution of Feedback from Individual Stars}

    \author{Brooke Polak
       \inst{1,2}\fnmsep\thanks{Fellow of the International Max Planck Research \newline School for Astronomy and Cosmic Physics at the University of Heidelberg}\orcidlink{0000-0001-5972-137X}
         \and
         Mordecai-Mark Mac Low\inst{2,3}\orcidlink{0000-0003-0064-4060}
         \and
        Ralf S. Klessen\inst{1,4}\orcidlink{0000-0002-0560-3172}
         \and
       Jia Wei Teh\inst{1}\orcidlink{0000-0001-7863-5047}
         \and
        Claude Cournoyer-Cloutier\inst{5}\orcidlink{0000-0002-6116-1014}
         \and
        Eric P. Andersson\inst{2}\orcidlink{0000-0003-3479-4606}
         \and
        Sabrina M. Appel \inst{6}\orcidlink{0000-0002-6593-3800}
         \and
        Aaron Tran\inst{7}\orcidlink{0000-0003-3483-4890}
         \and
        Sean C. Lewis\inst{3}\orcidlink{0000-0003-4866-9136}
         \and
        Maite J. C. Wilhelm\inst{8}\orcidlink{0000-0002-3001-9461}
         \and
        Simon Portegies Zwart\inst{8}\orcidlink{0000-0001-5839-0302}
         \and
        Simon C. O. Glover\inst{1}\orcidlink{0000-0001-6708-1317}
         \and
        \response{Steven Rieder\inst{9}\orcidlink{0000-0003-3688-5798}}
         \and
        Long Wang\inst{10}\orcidlink{0000-0001-8713-0366}
        \and
        Stephen L. W. McMillan\inst{3}\orcidlink{0000-0001-9104-9675}
       }

\institute{Universit\"{a}t Heidelberg, Zentrum f\"{u}r Astronomie, Institut f\"{u}r Theoretische Astrophysik, Heidelberg, Germany\\
           %\email{brooke.polak@uni-heidelberg.de}
           \email{bpolak@amnh.org}
        \and
         Department of Astrophysics, American Museum of Natural History, New York, NY, USA
         \and
         Department of Physics, Drexel University, Philadelphia, PA, USA
         \and
        Universit\"{a}t Heidelberg, Interdisziplin\"{a}res Zentrum f\"{u}r Wissenschaftliches Rechnen, Heidelberg, Germany
         \and
       Department of Physics and Astronomy, McMaster University, Hamilton, ON, Canada
         \and
        Department of Physics and Astronomy, Rutgers University, Piscataway, NJ, USA
         \and
        Department of Physics, University of Wisconsin--Madison, Madison, WI, USA
         \and
        Sterrewacht Leiden, Leiden University, Leiden, the Netherlands
         \and
        \response{Institute of Astronomy, KU Leuven, Celestijnenlaan 200D, B-3001 Leuven, Belgium}
         \and
        School of Physics and Astronomy, Sun Yat-sen University, Daxue Road, Zhuhai, 519082, China   
        }

   \date{Received December 4, 2023; accepted June 5, 2024}

% \abstract{}{}{}{}{} 
% 5 {} token are mandatory
 
    \abstract{  
    The mode of star formation that results in the formation of globular clusters and young massive clusters is difficult to constrain through observations. We present models of massive star cluster formation using the \textsc{torch} framework, which uses the Astrophysical MUltipurpose Software Environment (AMUSE) to couple distinct multi-physics codes that handle star formation, stellar evolution and dynamics, radiative transfer, and magnetohydrodynamics.
    We upgraded \textsc{torch} by implementing the N-body code \textsc{petar}, thereby enabling \textsc{torch} to handle massive clusters forming from $10^6\rm\, M_\odot$ clouds with $\ge10^5$ individual stars.  
    We present results from \textsc{torch} simulations of star clusters forming from $10^4,\, 10^5,\rm\, and\, 10^6\, M_\odot$ turbulent spherical gas clouds (named M4, M5, M6) of radius $R=11.7\rm\, pc$.
    We find that star formation is highly efficient and becomes more so at a higher cloud mass and surface density. 
    For M4, M5, and M6 with initial surface densities $2.325\times 10^{1,2,3}\rm\, M_\odot\, pc^{-2}$, after a free-fall time of $t_{\rm ff}=6.7,2.1,0.67\rm\, Myr$, we find that $\sim$30\%, 40\%, and 60\% of the cloud mass has formed into stars, respectively.
    \response{The end of simulation-integrated star formation efficiencies for M4, M5, and M6 are $\epsilon_\star=M_\star/M_{\rm cloud}=$36\%, 65\%, and 85\%.
    Observations of nearby clusters similar in mass and size to M4 have instantaneous star formation efficiencies of $\epsilon_{\rm inst}\leq30\%$, which is slightly lower than the integrated star formation efficiency of M4.}
    The M5 and M6 models represent a different regime of cluster formation that is more appropriate for the conditions in starburst galaxies and gas-rich galaxies at high redshift, and that leads to a significantly higher efficiency of star formation.
    We argue that young massive clusters build up through short efficient bursts of star formation in regions that are sufficiently dense ($\Sigma \ge 10^2 \rm\, M_\odot\ pc^{-2}$) and massive ($M_{\rm cloud} \ge 10^5\rm\, M_\odot$).
    In such environments, \response{stellar feedback from winds and radiation is not strong enough to counteract the gravity from gas and stars until a majority of the gas has formed into stars.}
    % the dynamical time of the cloud becomes short enough that stellar feedback cannot act quickly enough to slow star formation.
    }


\maketitle
%
%-------------------------------------------------------------------

\section{Introduction}

Globular clusters (GCs), which are found in every massive galaxy, are some of the most ancient objects in the Universe. They serve as fossils that can reveal the elusive environment and physics of the early phases of galaxy assembly \citep{Brodie2006ARA&A..44..193B,PortegiesZwart2010ARA&A..48..431P,Renaud2017MNRAS.465.3622R,Krumholz2019ARA&A..57..227K,Adamo2020SSRv..216...69A}. Yet because of their age, many aspects of cluster formation and evolution at high redshift are challenging to constrain through observation, and little is known about the efficiency and timescale at which gas is converted into stars to create such massive bound clusters. 

Though the progenitors of GCs are too old to observe, there are younger star clusters that are as massive as GCs and currently forming in nearby galaxies. These young massive clusters (YMCs) have masses  $M \ge 10^4 \, \rm M_\odot$ and ages  $<100$ Myr \citep{PortegiesZwart2010ARA&A..48..431P}. The discovery of these objects has indicated that the mode of extreme star formation that forms massive star clusters still occurs today. Notably, even more of these clusters are being discovered with JWST, as many YMCs in starburst galaxies are too embedded to have been seen by {\em Hubble} \citep{Whitmore2023ApJ...944L..14W}. Although it has been suggested that YMCs are the present day analogs to young GCs, this is debated in the literature (see \citealt{Renaud2020IAUS..351...40R}). 

Theory suggests that, despite the abundance of GCs, $\le 1\%$ of clusters survive to become GCs \citep{Fall2001ApJ...561..751F,Fall2005ApJ...631L.133F,Fall2006ApJ...652.1129F}. The conditions that lead to bound star clusters as massive as GCs remain a mystery, and observations of forming YMCs are sparse. Star formation must be fast and efficient enough to form bound stars that can survive the epoch of stellar feedback and the dispersal of the natal gas \citep{Lada2003ARA&A..41...57L}. The plethora of GCs suggests these conditions were very common in the early Universe.

The process of star formation in a cluster begins with the global gravo-turbulent collapse of giant molecular clouds \citep[GMCs;][]{Larson1981MNRAS.194..809L}. As the collapse proceeds, fragmentation creates overdense clumps within the GMC that begin to form stars \citep{MacLowRevModPhys.76.125,Mckeedoi:10.1146/annurev.astro.45.051806.110602, Klessen2016SAAS...43...85K}. The feedback from these stars, in the form of stellar winds, jets, and radiation, begins to clear out dense gas in and around the forming sub-clusters, slowing down the local (sub-cluster scale) and global (cloud scale) star formation rate \citep[SFR; e.g.,][]{Girichidis2020SSRv..216...68G,Lewis2023ApJ...944..211L}. Eventually, massive stars explode as supernovae (SNe), further dispersing gas. However, it has been argued that the efficiency at which stellar feedback slows global star formation diminishes with higher gas surface density \citep{Grudic2018MNRAS.475.3511G}. The sub-clusters eventually merge if they are mutually gravitationally bound, forming a final cluster cleared of all natal gas \citep{Krause2020SSRv..216...64K}. 

Many details of star cluster formation remain poorly understood due to the difficulty of modelling such a complex process. Stellar evolution and binary dynamics need to be resolved on timescales of years and distance scales of an AU, while the magnetohydrodynamics (MHD) of the collapsing gas covers regions several parsecs across, with crossing times of thousands to millions of years. Because of this, most computational star cluster formation models are limited and must make considerable approximations. \response{Many simulations do not form individual stars: some apply stellar feedback as a combined source in the center of the cloud \citep{Dale2005MNRAS.358..291D,Rahner2019MNRAS.483.2547R}, and others use sink particles representing sub-clusters \cite[e.g.,][]{Bate1995MNRAS.277..362B,Federrath2010} or extract the properties and feedback of individual stars from the sink particles \citep[e.g.,][]{Sormani2017,10.1093/mnras/stx1363,Kim_2017,10.1093/mnras/sty035,10.1093/mnras/sty1928}.} Other simulations do form single stars, but they do not resolve the stellar feedback of each individual star particle \citep{10.1093/mnras/stt1409,10.1093/mnras/stz1271}, instead including feedback from just the sink particles that created the stars. \response{Simulations of dwarf galaxies can capture star cluster mass functions and formation times, but they do so without collisional dynamics of star particles and are therefore unable to accurately capture dynamical properties such as velocity dispersion and size \citep{Lahen2019ApJ...879L..18L,Lahen2024MNRAS.530..645L,Andersson2024A&A...681A..28A}.}

Modelling individual stars is important, as this can change the efficiency and location of stellar feedback injection. Dynamical processes often eject high-mass stars \citep{Fujii2011Sci...334.1380F,Fujii2022b}, and the location of massive stars directly affects how and when gas is dispersed. Gas dispersal stops star formation. Models of sub-cluster feedback may overestimate the strength of feedback, as they do not allow for spatial separation between the stars in the sub-cluster. This lack of separation also changes the morphology of the gas, affecting the number of low-density channels in the gas that can vent thermal energy from the sub-cluster. The degree to which the sub-cluster and star-by-star approaches differ must be constrained.

There are a few models that do evolve individual stars with both stellar feedback and higher order gravitational dynamics \response{\citep{2020Wall,2021MNRAS.506.2199G,Fujii2021PASJ...73.1074F,Fujii2022a,Fujii2022b,2021Cournoyer-Cloutier,Lewis2023ApJ...944..211L,Wilhelm2023MNRAS.520.5331W,Cournoyer-Cloutier2023MNRAS.521.1338C}. While these models} include most of the relevant physics, they lack the computational efficiency to simulate star clusters forming from clouds of masses $>10^5$ M$_\odot$, and instead the models focus on simulating star clusters forming from low-mass clouds $\le10^5$ M$_\odot$. This leaves a sizeable gap compared to the observed mass range of GMCs. While clusters with mass $<10^5\rm\, M_\odot$ are comparable to Local Group observations, YMC and GC formation is out of their reach. Furthermore, most star formation takes place in GMCs of mass $\ge10^5$ M$_\odot$ \citep{McKee_1997,Murray_2010}. 

The goal of this work is to model the formation of massive clusters from their initial GMCs while following the formation of individual stars and their feedback. We aim to answer how and in what conditions YMCs can form while remaining bound throughout the onset of gas expulsion. We also seek to understand how efficient the process of star formation is in a cluster, what the timescale is over which star formation occurs, and whether the clusters formed from these massive clouds survive and remain bound or quickly disperse. We plan to compare our results to those that use a sub-cluster formation and feedback model.

To do this, we used the \textsc{torch} framework \citep{wall2019,2020Wall}. \textsc{torch} employs the Astrophysical Multipurpose Software Environment (AMUSE) framework to couple separate physics codes that handle MHD, radiative transfer, stellar evolution, and N-body dynamics. \textsc{torch} uses the MHD code \textsc{flash} \citep{flash,dubey2014}, which accounts for the evolution of the gas and the formation of sink particles and stars. Stellar feedback in the form of winds and SNe is included, and the effect of ionizing and non-ionizing radiation is followed using a ray-tracing approach \citep{FERVENT10.1093/mnras/stv1906}. The star formation model samples the \citet{2002Sci...295...82Kroupa}  initial mass function (IMF) to form stars from sink mass reservoirs \citep{wall2019}. \textsc{SeBa} \citep{SeBa} tracks stellar evolution from the zero-age main sequence, and, in the original version of \textsc{torch}, \textsc{ph4} \citep{ph4} + \textsc{multiples} \citep{amusebook} handled the stellar dynamics.

In that version \citep{wall2019,2020Wall}, \textsc{torch}  could not handle the hundreds of thousands of stars that form in massive GMCs $>10^5\rm\, M_\odot$. In this work, we solve this problem by making three updates: 1) We replace the combination of the N-body code \textsc{ph4} and the higher-order interactions \textsc{multiples} with the code \textsc{petar} \citep{PeTar}; 2) we agglomerate stars with masses $< 4\rm\, M_\odot$ into summed-mass dynamic star particles with masses of $\geq 4\rm\, M_\odot$; and 3) we mass load the stellar winds to reduce the peak temperatures beyond their termination shocks. These modifications enabled \textsc{torch} to then model clouds with an initial mass of up to $10^6\rm\, M_\odot$ that form hundreds of thousands of individual stars. 

We present simulations of star clusters forming from turbulent spherical clouds with masses of $10^4,\, 10^5$, and $10^6 \rm\, M_\odot$. Each of these clouds is almost identical in terms of their initial properties, with only mass and density scaled between them. Our study investigates whether the formation of YMCs parallels that of low-mass clusters or if it varies significantly with initial cloud mass and density. 

This paper is the first in a series exploring the results of these simulations. In this paper, we describe the \textsc{torch} code, the new features integrated into \textsc{torch} for handling massive GMCs, and the initial conditions of our three clouds in Sect.~\ref{sec:methods}. We analyze the time evolution of global gas and stellar properties in Sect.~\ref{sec:results}. In Sect.~\ref{sec:discussion}, we discuss the results of our analysis, and in Sect.~\ref{sec:conclusions} we conclude with a summary of the most important results. \response{We provide a data repository containing a sampling of the simulation data corresponding to the panels in Figure~\ref{fig:time_panel}
% and the analysis scripts used to extract the global properties of the clusters. We also include 
, the HTML file of the three-dimensional interactive plot shown in Figure~\ref{fig:plotly}, and the code used to generate the interactive plot.} %This repository is available here:}

\section{Methods} \label{sec:methods}

\subsection{Standard \textsc{torch}} \label{sec:Torch}

\response{\textsc{torch}\footnote{\textsc{torch} version used for this work: \href{https://bitbucket.org/torch-sf/torch/commits/tag/massive-cluster-1.0}{https://bitbucket.org/torch-sf/torch/commits/tag/massive-cluster-1.0}}} is built upon the AMUSE framework, which couples multiple autonomous astrophysical codes. We chose codes that allowed efficient calculation of the disparate physical processes at work in star cluster formation. 

The \textsc{torch} framework incorporates the adaptive mesh refinement MHD code \textsc{flash} v4.6.2 \citep{flash,dubey2014} with a number of enhancements to the base version of \textsc{flash}. The base \textsc{flash} handles the MHD and sink particle formation and evolution. The modifications to \textsc{flash} presented in \citet{wall2019,2020Wall} include heating and cooling, ionization, radiation transfer \citep[using ray-tracing; see][]{FERVENT10.1093/mnras/stv1906}, and feedback injection from stars. Stellar feedback is implemented in \textsc{flash} in the form of ionizing extreme ultraviolet (EUV) and non-ionizing far ultraviolet (FUV) radiation in the form of radiative heating and radiation pressure, as well as mechanical feedback from stellar winds and SNe. \response{FUV rays are terminated when their flux drops below $F_{\rm ray} \leq 16.9~G_0~e^{-3.5~A_v}$, where $A_v$ is the visual extinction and $G_0$ is the Habing flux. This cutoff is $10\times$ the applied background FUV field of $F_{ext}=1.69~G_0~e^{-3.5~A_v}$ \citep{Draine1978ApJS...36..595D}. This limits the number of low energy rays on the grid for computational efficiency.} We used the HLLD Riemann solver \citep{Miyoshi2005JCoPh.208..315M} in \textsc{flash} paired with third-order piecewise parabolic method reconstruction \citep{Colella1984JCoPh..54..174C}. 

To avoid artificial fragmentation, the Jeans length, 
\begin{equation}
\lambda_{\rm J}=\sqrt{\pi c_s^2 / (G\rho)} \label{eq:jeans},
\end{equation}must be resolved by at least four cells \citep{Truelove1997ApJ...489L.179T}. \response{We used a refinement criterion of 12 cells per Jeans length along with a derefinement criterion of 24 cells per Jeans length.} As density increases during collapse, the Jeans length decreases until this criterion is no longer met at the highest level of AMR refinement. Sink particles were \response{used} to collect the gas that \response{exceeds} this density. \response{The Truelove criterion} sets the sink radius to $R_{\rm sink}=2.5\Delta \rm x_{min}$ and gives the sink threshold density during the entire run as
\begin{equation}
    \rho_{\rm sink}=\frac{\pi c_s^2}{G \lambda_{\rm J}^2}=\frac{\pi c_s^2}{G (5\Delta \rm x_{min})^2} \, ,
\end{equation}
where $c_{\rm s}$ was evaluated using the initial temperature of the gas. \response{(If the gas heats during the run, the dense gas will be better resolved, making this a worst-case limit for the required density resolution.)}

On each time step, the mass of gas within a distance $R_{\rm sink}$ of a sink particle that satisfies the criteria outlined in \citet{Federrath_2010} is added to that sink's mass reservoir for creating stars. When a sink forms, it randomly samples the Kroupa IMF \citep{2002Sci...295...82Kroupa} and stores a \response{long} list of \response{potential} star masses to form \cite[see also][]{Sormani2017}. Each time step, the sink forms \response{as many} stars from this mass list \response{as possible} until its \response{current} mass reservoir is depleted. It again forms one or more stars the next time it has accreted enough mass for at least the next star on the list. This is the standard stellar mass sampling method used in \textsc{torch} \citep{wall2019}. Star positions are randomly sampled from a uniform spherical distribution within the sink's accretion radius. Star velocities are set by the sink velocity \response{added to} an additional isotropic velocity dispersion with a Gaussian distribution \response{having a} standard deviation of the local sound speed. 

Star particles are initially formed as zero-age main sequence stars, neglecting pre-main sequence evolution. Subsequent stellar evolution is tracked with \textsc{SeBa} \citep{SeBa}, which passes the evolutionary properties informing stellar feedback to \textsc{flash}. The N-body dynamics of the stars are calculated using \textsc{petar} \citep{PeTar}, which is discussed further in the next section. Stars dynamically interact with the gas in \textsc{flash} through the AMUSE hierarchical coupling \citep{PORTEGIESZWART2009369amuse1} based on the gravity-bridge algorithm of \citet{Fujii2007PASJ...59.1095F}.

\subsection{\textsc{petar} N-body} \label{sec:PeTar}

\begin{figure}
\centering
    \includegraphics[width=\hsize]{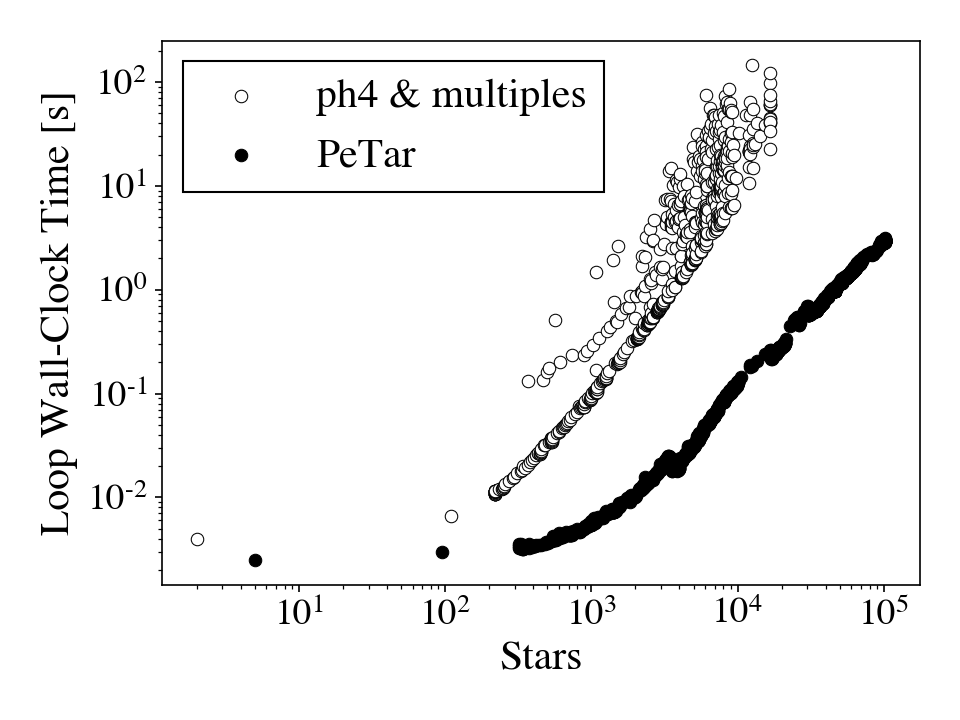}
    \caption{Wall-clock time for an evolution step for \textsc{petar} and \textsc{ph4}+\textsc{multiples} given the number of stars. }
    \label{fig:petarvsph4}
\end{figure}

\textsc{torch} was first designed to use the N-body code \textsc{ph4} \citep{ph4} to handle direct stellar dynamics, paired with \textsc{multiples} \citep{amusebook} to track binary orbital evolution and higher order perturbations. For \textsc{torch} runs using an initial cloud of $10^4$ M$_\odot$ and producing only a few thousand stars, this works well. However, the computational cost becomes unfeasible when pushing to higher initial cloud masses that produce far more than $10^4$ stars with over a few hundred binary systems. This is because \textsc{multiples} is a serial Python code, so with many interacting stars computation times become impractical. To speed up \textsc{torch}, we replaced \textsc{ph4} and \textsc{multiples} with \textsc{petar} \citep{PeTar}.

\textsc{petar} is a state-of-the-art gravitational dynamics code optimized for solving the stellar dynamics of systems with millions of stars. It accomplishes this by dividing gravitational interactions into three regimes: distant interactions calculated with a \citet{BarnesHut1986Natur.324..446B} tree and handled by the framework for developing parallel particle simulation codes \citep[\textsc{fdps};][]{Iwasawa2016,Iwasawa2020}, nearby interactions solved with a fourth-order Hermite direct N-body integrator \citep{Makino1992PASJ...44..141M}, and close interactions (binaries and higher order systems/perturbations) solved using the Slow Down Algorithmic Regularization (SDAR) technique \citep{SDAR2020MNRAS.493.3398W}. For each particle, the force from neighboring particles is solved depending on what distance regime they are in, with a mass-dependent factor to increase the distance over which massive particles are considered close neighbors. The SDAR feature for handling higher-order dynamics is the novel component of \textsc{petar}, enabling it to handle large numbers of binaries and higher order systems in parallel. 

In Figure~\ref{fig:petarvsph4}, we plot the wall-clock time per evolution step for each of the N-body codes considered. For reproducibility, this test was done with the parameters $r_{out}=0.001\rm\, pc$, $r_{bin}=100\rm\, AU$ for \textsc{petar} and the stellar interaction radius $r_{\rm int}=15\rm\, R_\odot$ for \textsc{multiples}. \textsc{petar} is significantly faster and consistently performs well as the number of stars increases. The variability in the performance of \textsc{ph4} and \textsc{multiples} is due to \textsc{multiples} taking longer if there are many third-body perturbations in a given step. We note that this test was done with single stars only; the scaling for a run with primordial binaries will be different.

When running \textsc{petar} in \textsc{torch}, the time step of the long distance particle tree must be set (\verb|dt_soft|), as well as the changeover radius between direct N-body and tree method for force calculations (\verb|r_out|). If the user sets these two parameters, all other parameters are set automatically. In \textsc{torch}, the MHD code sets the global time step for all worker codes based on the Courant condition. The tree time step was set as the nearest power of two in code units below the requested time step, as a power of two is required by \textsc{petar} (like most N-body codes). This sets \verb|dt_soft|. \response{For M5 and M6 w}e set the outer radius \verb|r_out| to $0.001 \rm\, pc$, the standard value used in \textsc{petar} simulations. We used a softening length of $\ell = 15\rm\, R_\odot$ \response{and a binary search radius of $r_{\rm bin}=16.5\rm\, AU$}. 
\response{For M4 we used a larger $r_{\rm out}=10~r_{\rm sink}=7.8\rm~pc$ to ensure accurate force calculations given the small number of stars and low stellar density. This corresponds to $r_{\rm bin}=0.63\rm~pc$.}

The code handling stellar mergers within \textsc{petar} is not active within the AMUSE framework, which results in star particles that approach within each others' softening radius and should merge instead ending up with identical positions, leading to a halt in code execution. We have implemented code to check for particles in this state, and merge them. \edit{We intended to use \textsc{SeBa} to update the stellar mass of the surviving star, but later testing revealed that the surviving star's mass remained unchanged. One star in the merger is removed meaning that stellar mass is unphysically lost. The effect of this error is negligible due to low merger rates: there are 0, 2, and 4 mergers in M4, M5, and M6, respectively. All of these mergers involve stars $<7\rm~M_\odot$. M5 and M6 lose only $8.4\rm~M_\odot$ and $22\rm~M_\odot$ of stellar mass due to unphysical mergers over the course of the simulations.} 

\subsection{Stellar modifications} \label{sec:sf_mods}

We made three alterations to the star formation and evolution procedures that vary from standard \textsc{torch} to accommodate the several orders of magnitude increase in number of stars formed when increasing the initial cloud mass from $10^4\rm\, M_\odot$ to $10^6\rm\, M_\odot$. 

\begin{enumerate}[I.]
    \item We agglomerated low-mass star particles below $M_{\rm agg}=4\rm\, M_\odot$ as they formed until their summed mass is $\ge M_{\rm agg}$. Then, a star particle is formed with a mass equal to the sum of the low-mass stars. This reduces the strain on the N-body calculations by reducing the number of dynamical star particles by $90\%$.
    \item We mass-loaded stellar winds to raise the Courant time step by limiting the temperature of wind-blown bubbles to $T_w=3\times 10^5\rm\, K$, which significantly sped up the simulations. This resulted in smaller, cooler, momentum-conserving bubbles instead of hot energy-conserving bubbles. The primary effect of wind feedback in cluster formation is to clear out extremely dense gas in order to allow ionizing radiation to form expanding \ion{H}{II} regions. In this dense gas even hot stellar wind bubbles cool quickly, so there is little change in behavior in this regime. 
    \item We only injected feedback from stars above $20\rm\, M_\odot$ to reduce the cost of ray-tracing. Massive stars output most of the ionizing radiation and mechanical wind energy in clusters: by setting this limit we lost less than $20\%$ of the total feedback energy. Stars below the feedback cutoff mass did not go SN within the time frame of our simulations ($\le 10\rm\, Myr$).
\end{enumerate}

We further explain and examine the effects of these modifications in Appendix \ref{app:sf_mods}, including providing a quantitative analysis of the amount of total energy lost by excluding feedback for stars $<20\rm\, M_\odot$ in the M6 model.

\subsection{Initial conditions} \label{sec:ics}

The initial properties of our three clouds are listed in Tables \ref{table:test} and \ref{table:control}. 
We chose to keep the radius of all three clouds constant at $R_{\rm cl}=11.7\ {\rm\,pc}$. The radius was kept the same to have the same spatial distribution of star formation for each run. 
Constant radius allows the cell resolution and size of sink particles to be the same between the three simulations, and it facilitates directly comparing the morphology and dynamics of the forming clusters. 

Consequently, the average initial densities of the clouds are 1.5, 15, and 150 ${\rm\, M}_\odot\mbox{ pc}^{-3}$, or $10^{-22},\ 10^{-21}$, and $10^{-20} \mbox{ g cm}^{-3}$. The column densities of these clouds are $2.325 \times 10^{1, 2, 3}$ M$_\odot$~pc$^{-2}$, respectively.  
Assuming a 9:1 number ratio of H:He, resulting in a mean molecular weight of $\mu=1.3$, this gives total particle number densities of $n=46,\ 460,\ \mbox{and } 4600\mbox{ cm}^{-3}$. Each cloud has a column density consistent with observations. Observations show a strong positive correlation between the mass and density of GMCs in PHANGS galaxies \citep{Sun2022AJ....164...43S}, suggesting mass and density should be varied together.

The initial clouds must be in pressure equilibrium with their surroundings to avoid unphysical shocks from pressure imbalances at their surfaces. The M4 and M5 clouds are in the pressure regime where there is a stable two-phase medium at solar metallicity and Milky Way background UV field \citep{field1969, wolfire2003}, meaning there is a set of temperatures for the cloud and background for a given cloud density where the cold dense cloud and the warm ambient medium are both in thermal equilibrium at equal pressure. The cloud temperatures for the M4 and M5 clouds are $T_{\rm cl}=103$~K and $28$~K, respectively, and the corresponding background temperatures and number densities are $T_{\rm amb}=9,000
\mbox{ and }4,000\mbox{ K}$, and $n_{\rm amb}=3\mbox{ and } 1\mbox{ cm}^{-3}$. The M6 cloud, however, is at a high enough pressure that a two-phase medium no longer exists. Only the cold phase can be in thermal equilibrium. This means that the low-density envelope of the M6 cloud is inherently not in thermal equilibrium.
To minimize the pressure imbalance with the core, we therefore raised the background density to $n_{\rm amb}=100\ {\rm\, cm^{-3}}$. Both the cloud and background medium for M6 are at a temperature of $T_{\rm cl} = T_{\rm amb}=50$~K. 

\begin{figure*}
\centering\includegraphics[width=0.8\textwidth]{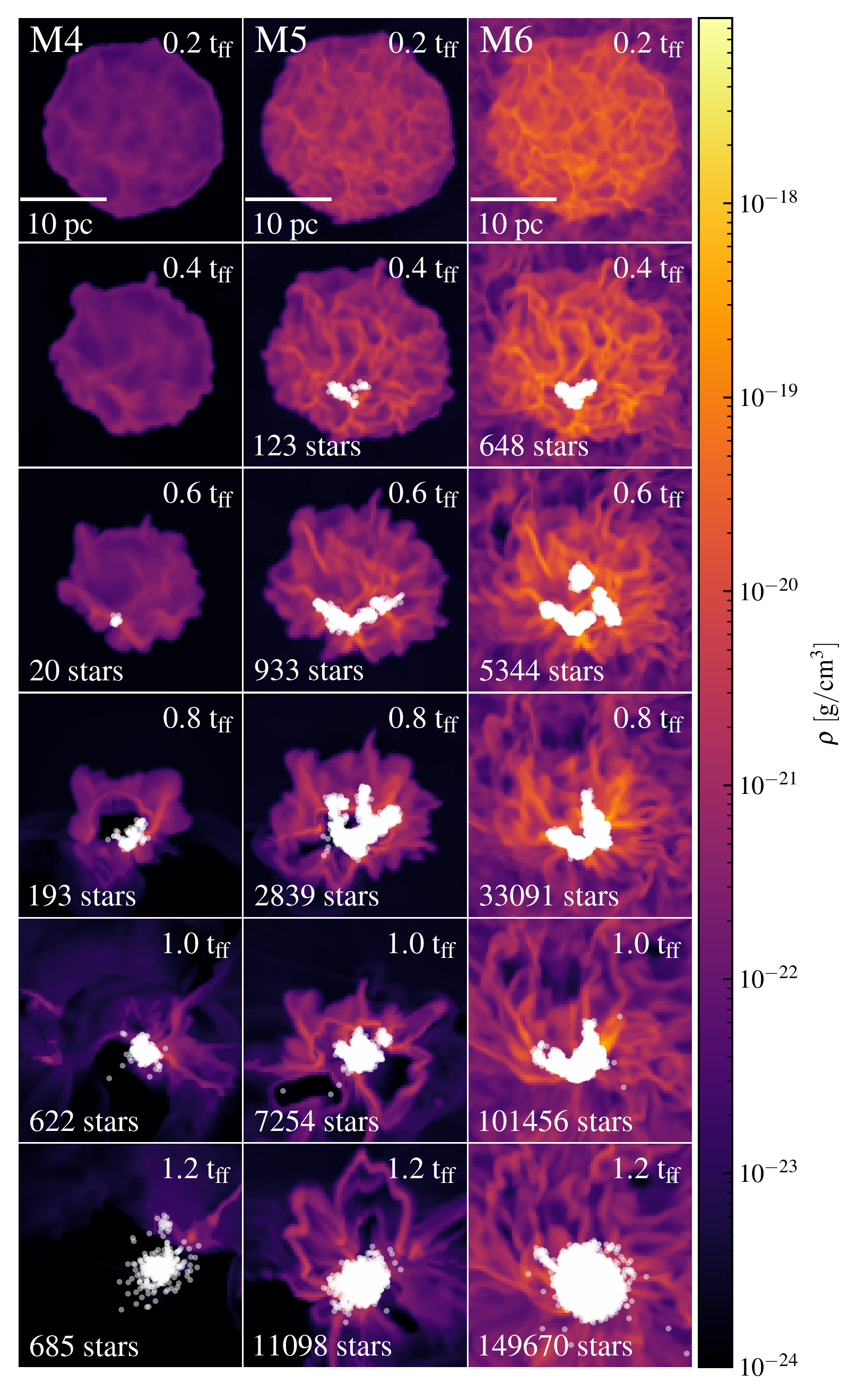}
    \caption{Slice plots of the three simulations in the $x$-$y$ plane over time. The plane of the slices for a given cloud is the center of stellar mass in the final snapshot. Stellar positions are shown by white dots. The free-fall times $t_{\rm ff}$ are given in Table~\ref{table:test}. The number of stars shows the amount of star particles in the domain, not the number sampled from the IMF. Due to our agglomeration of low-mass star particles, the number of stars sampled from the IMF is $\sim10\times$ greater.
    }
    \label{fig:time_panel}
\end{figure*}

\begin{table}
\caption{Model parameters.}   
\label{table:test}   
\centering   
\begin{tabular}{l|ccc}          
\hline\hline                        
Parameter & M4 & M5 & M6 \\   
\hline                                   
    $M_\mathrm{cl}$ [M$_\odot$] & $10^4$ & $10^5$ & $10^6$ \\
    $\rho_{c},\Bar{\rho}$ [M$_\odot$ pc$^{-3}$] & 2.8, 1.5 & 28, 15 & 280, 150 \\
    $\Sigma\ {\rm [M_\odot pc^{-2}]}$ & 23.25 & 232.5 & 2325 \\
    $\lambda_{\rm J}$ [pc] & 10.0 & 3.2 & 1.0 \\
    $t_\mathrm{ff}$ [Myr] & 6.7 & 2.1 & 0.67 \\ 
\hline  
\end{tabular}
\tablefoot{Rows: Cloud mass, cloud central and average volume density, initial column density, Jeans length (Eq. \ref{eq:jeans}) at initial temperature, and free-fall time.}
\end{table}

\begin{table}
\caption{Control parameters.}   
\label{table:control}   
\centering   
\begin{tabular}{l|cc}          
\hline\hline                        
Parameter & Value & Units \\   
\hline                                   
    $R_\mathrm{cloud}$ & 11.7 & pc\\
    $R_\mathrm{box}$ & 20.0 & pc\\ 
    $\alpha_v$ & 0.15 & -\\
    $\Delta$x$_\mathrm{min}$ & 0.3125 & pc\\
    $\Delta$x$_\mathrm{max}$ & 1.25 & pc\\
    $R_\mathrm{sink}$ & 0.78125 & pc\\
    $\rho_\mathrm{sink}$ & $8\times 10^{-21}$ & g cm$^{-3}$\\ 
    $M_\mathrm{sink}$ & 246 & M$_\odot$\\ 
    $M_\mathrm{feedback}$ & 20 & M$_\odot$\\ 
    $M_\mathrm{n-body}$ & 4 & M$_\odot$\\ 
    $M_\mathrm{IMF}$ & 0.08--100 & M$_\odot$\\ 
\hline  
\end{tabular}
\tablefoot{Rows: Radius of cloud, half-width of box, virial parameter, minimum cell width, maximum cell width, sink radius, sink threshold density, initial sink mass, minimum feedback star mass, agglomeration mass of low-mass stars, mass sampling range of IMF.}
\end{table}

The initial conditions described in the rest of this section and summarized in Table~\ref{table:control} apply to all three clouds. The clouds have a Gaussian density profile \citep{Bate1995MNRAS.277..362B,Goodwin2004A&A...414..633G} with the standard deviation set such that the ratio of the cloud's central to edge density is 3:1. The simulation domain is a cube of half-width $R_{\rm box}=20~\rm\, pc$ with outflow boundary conditions. The outflow boundaries do allow gas flow onto the grid from ghost zones if the velocity at the edge of the grid is directed inward. \response{The inflow of gas from the boundary is minimal: more gas exits the simulation than enters in all of our runs. Inflow was allowed, though, to prevent vacuums from forming at the boundaries.}
We used three refinement levels, yielding cell sizes that range from $\Delta x_{\rm min}=0.3125$~pc to $\Delta x_{\rm max}=1.25$~pc. 
Refinement and derefinement of the grid was determined by the Jeans criterion described in Sect.~\ref{sec:Torch} and based on temperature and pressure gradients. \response{The latter trigger refinement when the adapted \citet{Lohner1987CMAME..61..323L} estimator\footnote{This is a modified second derivative which is normalized by the average of the gradient over a computational cell.} of temperature or pressure exceeds 0.98 and trigger derefinement when the estimators drop below 0.6. We are interested in global formation properties of clusters rather than the fragmentation of the cloud or the origin of the IMF, so we find the chosen resolution to be sufficient.}

We initially imposed a \cite{Kolmogorov1941DoSSR..30..301K} turbulent velocity spectrum on all the gas in the domain. \response{The peak Mach numbers for the turbulent spectrum are $\mbox{Ma}=30.3$ for M6, $\mbox{Ma}=12.9$ for M5, and $\mbox{Ma}=2.1$ for M4.} The same random seed was used to generate the turbulent velocity spectrum for all three clouds. This ensured the same turbulent collapse patterns, minimizing differences in the formation, location, and morphology of dense cores. From the edge of the cloud to the domain boundary, we linearly tapered the magnitude of the turbulent velocities from $100\%$ to $25\%$. This tapering does not affect the low-density ambient background of the M4 and M5 cloud, but helps with stability in the M6 cloud by mixing the border of the cloud, where there is a small pressure jump. 

The sink accretion radius and threshold density, derived in Sect. ~\ref{sec:Torch}, are $r_{\rm sink}= 2.5\Delta x_{\rm min} = 0.78$ pc and $\rho_{\rm sink}=8.35\times 10^{-21} \mbox{ g cm}^{-3}$. This gives an initial sink mass resolution of $m_{\rm sink}=245\,M_\odot$, meaning that when a sink initially forms it will accrete and form approximately $m_{\rm sink}$ worth of stellar mass, given the sink's threshold density and accretion radius. The IMF sampling mass range is 0.08--100 M$_\odot$. The lower end is the hydrogen-burning limit, while the upper end is the most massive star thought to form in a star cluster with stellar mass $\mathrm{\approx10^4\,M_\odot}$ \citep{Weidner10.1111/j.1365-2966.2009.15633.x}. \response{This is the expected stellar mass limit for a cluster similar to M4, so we chose this value as a fixed parameter for consistency between the three clusters.}

The critical virial ratio for stability is $\alpha_v=E_{\rm kin}/|E_{\rm pot}| =0.5$, below which collapse occurs. 
Massive clouds tend to be sub-virial, with clouds of $~10^6\ \mathrm{M}_\odot$ observed to have virial parameters of $\alpha_v\approx 0.05-0.35$ \citep{2013ApJ...779..185K}, though some surveys see super-virial massive clouds \citep[see Fig. 2 of ][]{Chevance2023ASPC..534....1C}. We note that these values have been converted from the different virial parameter definition in \cite{2013ApJ...779..185K}. Therefore, we chose a fiducial virial parameter value of $\alpha_v=0.15$ for our models to promote rapid onset of collapse.

Magnetic fields are prevalent in the interstellar medium \citep{Crutcher2003LNP...614..155C} and affect the collapse of GMCs and subsequent star formation. Although they are not the dominant factor in determining how star formation proceeds within a cloud, their presence has been shown to alter the fragmentation of cores \citep{Price2008MNRAS.385.1820P,Peters2011ApJ...729...72P} and slow down the global evolution of the cloud \citep{Heitsch2001ApJ...547..280H}.
With a strong enough field, clouds can be supported against gravitational collapse \citep{Heiles76}, although generally observed magnetic fields are not strong enough to inhibit collapse \citep{Klessen2016SAAS...43...85K}. 
The critical value of the mass-to-flux ratio for a cloud to be supported by magnetic fields against gravitational collapse is given by \citep{1976Mouschovias,Mouschovias1991}
\begin{equation} \label{eq:MPhicrit}
    \Bigg[\frac{M}{\Phi}\Bigg]_c=\frac{\zeta}{3\pi}\sqrt{\frac{5}{G}}=490\ \mathrm{\frac{g}{Gauss\ cm^2}}\;,
\end{equation}
where $G$ is the gravitational constant and a correction factor $\zeta=0.53$ for a uniform sphere is used \citep{1966Strittmatter}.

In our simulations, each cloud's initial magnetic field $\vec{B} = B_z \hat{z}$ is uniform in $z$ and decreases radially in the $x$-$y$ plane, following the mid-plane density $\rho(x,y,z=0)$ as
\begin{equation}
    B_z(x,y) = B_0 \exp\left[ - (x^2+y^2) \ln(3) / R_\mathrm{cl}^2 \right],
\end{equation}
with $B_0 = 0.185,1.85,18.5\,\mathrm{\mu G}$ for the M4, M5, and M6 clouds, respectively. These values match observations for M5 and M6, while the field is a factor 10 weaker for M4 \citep{Crutcher2010ApJ...725..466C}.
The integrated magnetic flux
$\Phi = 2\pi B_0 R_\mathrm{cl}^2 / (3\ln(3))$,
so all clouds have an initial mass-to-flux ratio
$M_\mathrm{cl}/\Phi = 4.5 \times 10^4 \,\mathrm{g\;Gauss^{-1}\;cm^{-2}}$
much larger than Equation~\eqref{eq:MPhicrit}. The initial magnetic fields are thus weak and do not inhibit collapse in any of our simulations.

\section{Results} \label{sec:results}

\subsection{Cluster formation overview} \label{subsec:overview}

At the onset of the simulation, each cloud begins to gravitationally collapse. Turbulent velocities fragment the cloud and create overdense hubs and filaments. Because the same random seed was used in all three clouds to generate the initial turbulent velocity spectrum, the web of dense gas is the same for each cloud. This means that the spatial distribution of star formation is similar for all three clouds. This can be seen in the time evolution of the three clouds in Figure~\ref{fig:time_panel}. The first stars all form in the largest over-density in the middle of the bottom of the cloud. Then, more stars form along the filaments of the dense cloud forming a V shape. The M5 and M6 clusters in particular look very similar in terms of sub-clustering and merging. The M4 cluster forms significantly fewer stars and therefore fewer sub-clusters. 

By a free-fall time $t_{\rm ff}$, the sub-clusters in M4 and M5 have mostly merged, forming a single central spherical cluster. The M6 model is still forming stars in various sub-clusters and has not assembled its main central cluster yet. By looking at the spatial distribution of sub-clusters and the density of the gas, one can see that stellar feedback becomes most efficient once the sub-clusters have merged into a single cluster. Whether feedback is only strong enough to disperse gas when clustered or this is coincidental with the timing of feedback needs further examination, but this is outside of the scope of this introductory paper. Low density bubbles begin to occupy a significant fraction of the cloud volume once the central star cluster has been assembled.

Once most of the stars have formed, the efficiency of stellar feedback for dispersing the natal gas varies greatly for the three cloud masses. The final row in Figure~\ref{fig:time_panel} shows the M4, M5, and M6 systems at $1.2 t_{\rm ff}$. At this point, only the M4 and M5 clusters have blown large bubbles. The feedback from the M6 cluster has hardly slowed the collapse of the densest gas, and rapid star formation continues. The M4 cloud has dispersed nearly all of the remaining gas, and star formation has halted completely. 

\subsection{Visualizing cluster morphology}

\begin{figure*}
\centering
    \includegraphics[width=0.9\textwidth]{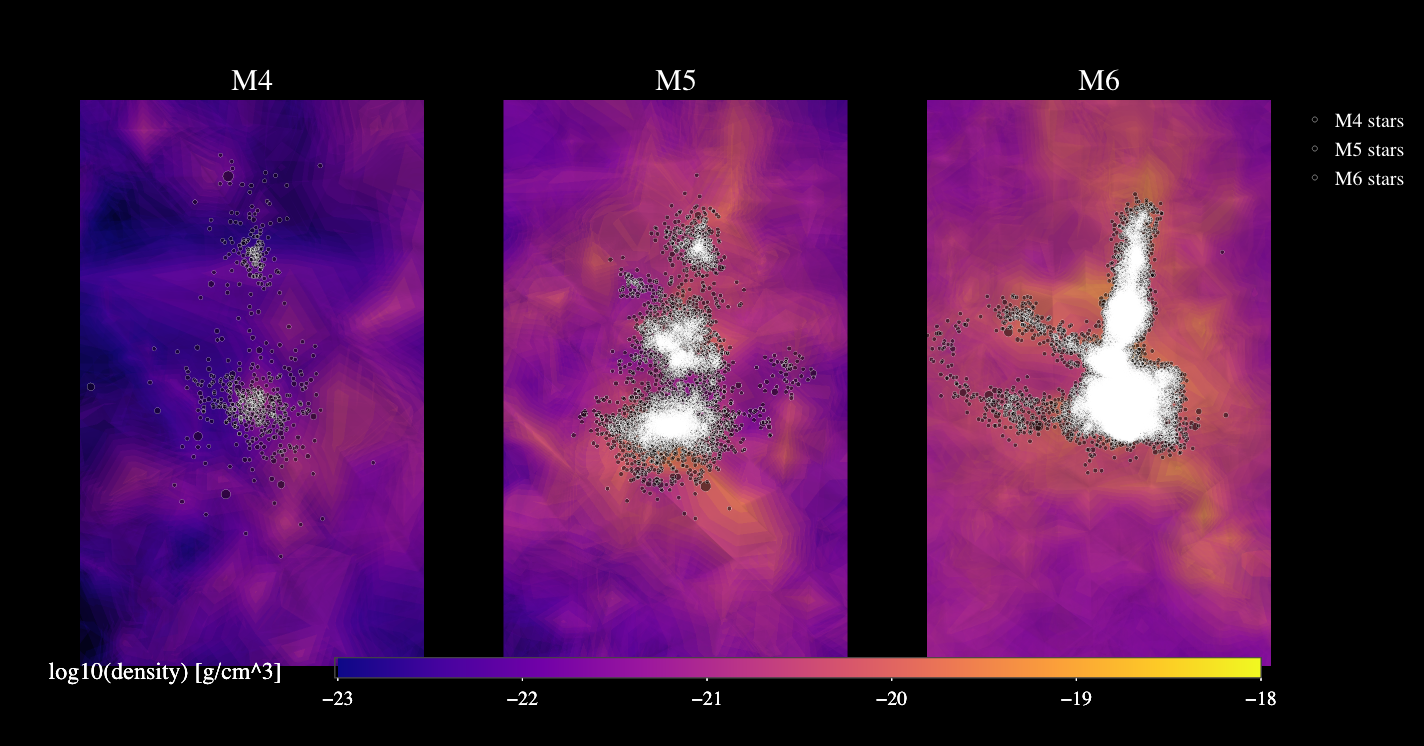}
    \caption{Still of the interactive plot of the embedded M4, M5, and M6 clusters (left to right) at $1\, t_{\rm ff}$. The interactive plot file is available for download \response{from the repository.}
    \label{fig:plotly}
    }
\end{figure*}

The complex 3D structure of star clusters is hard to visualize using 2D plots. Figure~\ref{fig:plotly} shows a still of an interactive plot of the M4, M5, and M6 simulations after one free-fall time generated with Plotly \citep{plotly}. After downloading the HTML file available in the online version of this paper, readers can zoom, pan, and rotate for a complete look at the morphology of each cluster. The color is an isosurface of the gas density in log scale, and the points are stars with sizes scaled to their stellar radius. This tool makes it clear just how non-spherical these clusters are. Comparing the still of the interactive plot to the slice plots in Figure~\ref{fig:time_panel}, one can already extract much more information on the system's morphology.

Zooming into the core of each cluster shows the immense stellar density of the M6 cluster, whereas in the M4 cluster one can easily distinguish individual stars. The gas is also far less dense in the M4 system compared to M5 and M6. 

The shape of the M6 cluster is highly irregular. Stemming from the largest cluster, one can see a row of sub-clusters forming along a filament. Branching perpendicularly off this filamentary cluster are two more star forming filaments in a configuration resembling the letter ``F.'' The M5 cluster has a shape  congruous to the shape of M6, but with fewer stars bridging the gaps between clusters in the main filament. The M5 cluster also has only one finger perpendicular to the main filament, which contains many fewer stars than the fingers of the M6 cluster. The M4 cluster is much less dense, with its few stars outlining the same core filament cluster seen in M5 and M6. However, in M4 sub-clusters can no longer be distinguished. The M4 sub-clusters already merged into a singular central cluster.

\begin{figure*}
\centering
    \includegraphics[width=1\textwidth]{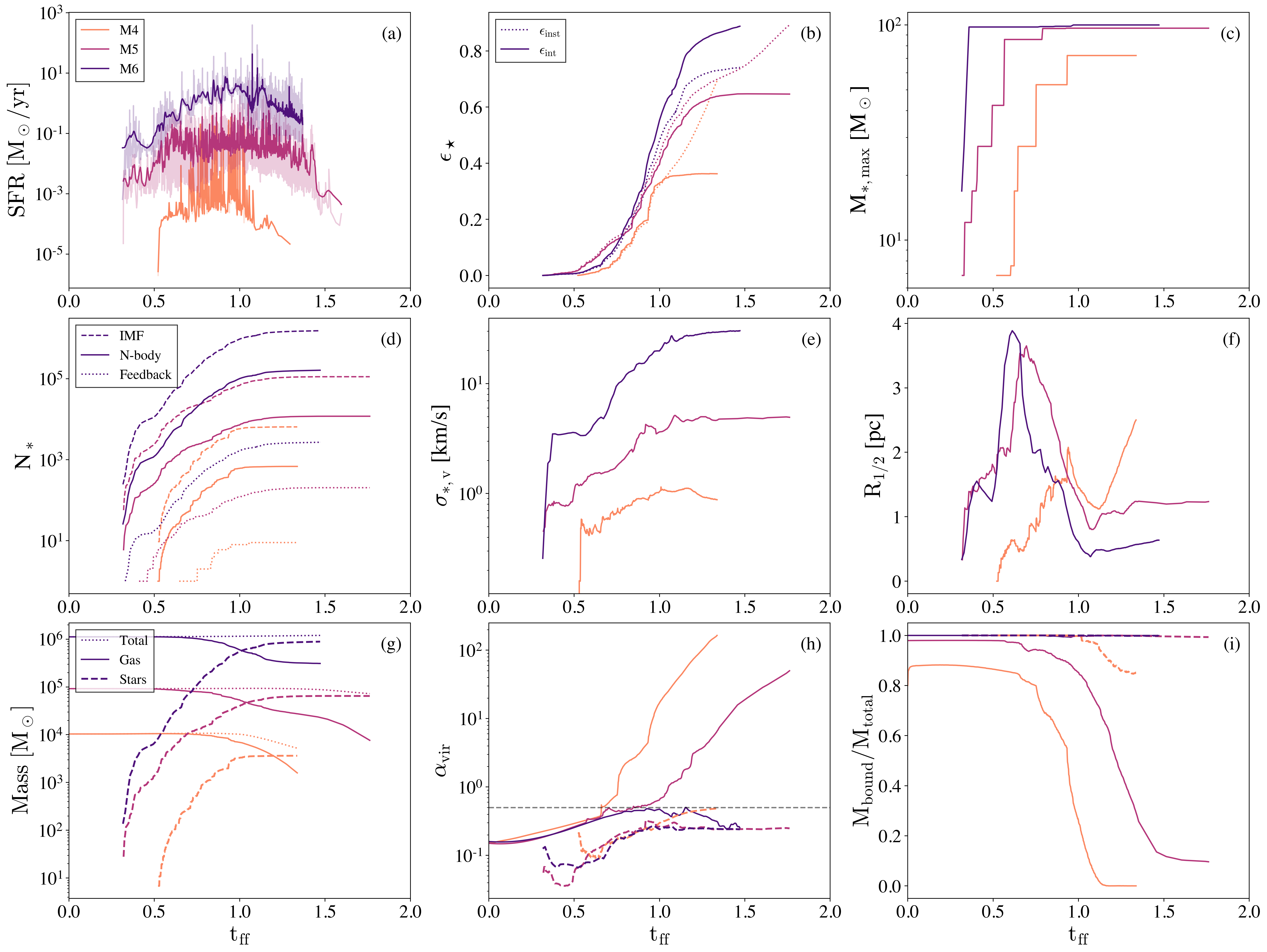}
    \caption{Global properties of the clusters and gas over time for models M4 {\em (orange)}, M5 {\em (maroon)}, and M6 {\em (blue-violet)} in units of free-fall time $t_{\rm ff}$ of the initial cloud (given in Table~\ref{table:test}).  
    From top left to bottom right: {\em(a)} SFR, where the transparent lines show the SFR at each star formation event, and the solid lines give the SFR smoothed using a Gaussian filter with $\sigma=0.005 t_{\rm ff}$. \response{{\em(b)} Instantaneous and integrated SFEs of the clouds, where $\epsilon_{\rm inst}=M_\star/(M_{\rm gas}+M_{\rm sink}+M_\star)$ and $\epsilon_{\rm int}=M_\star/M_{\rm cloud}=\epsilon_\star$.} {\em(c)} Most massive star formed. {\em(d)} Number of formed stars. {\em Dashed line:} actual number of stars that would form from sampling the IMF given the amount of gas mass collected for star formation by sink particles. {\em Solid line:} number of stars followed in \textsc{torch} after the sampled stellar population below $4\, M_\odot$ has been agglomerated. {\em Dotted line:} number of stars above $20\, M_\odot$ on the grid that are generating feedback. The number of stars can drop due to SN, mass loss, or exiting the grid. {\em(e)} 3D stellar velocity dispersion. {\em(f)} \response{Half-mass radius of the entire star cluster.} {\em(g)} \response{Total mass {\em (dotted line)}, mass of stars {\em (dashed line)} and gas {\em (solid line)} on the grid.} {\em(h)} Virial parameter of stars {\em (dashed line)} and gas {\em (solid line)}, where $\alpha_v=0.5$ is the equilibrium value. {\em(i)} Fraction of mass bound for stars {\em (dotted line)} and gas {\em (solid line)}.
    }
    \label{fig:global_ff}
\end{figure*}

\subsection{Star formation history}
\label{subsec:SFH}
The global properties of the star clusters that form from the M4, M5, and M6 clouds over the period of star formation are shown in Figure~\ref{fig:global_ff} as a function of time in units of the global free-fall time of the cloud $t_{\rm ff}$ (see Table~\ref{table:test}). For reference, this same figure is shown as a function of physical time in Appendix \ref{app:props_physical}. We analyze these properties and discuss how they compare across the three clouds to assess the effect of the initial cloud mass and density on the resultant cluster properties. Table \ref{table:results} highlights the final properties of the clusters formed in the three models.

Our results suggest that star formation in a GMC is a fast and efficient process regardless of initial cloud mass and density, with all three clouds converting at least $30\%$ of their initial gas into stars within an initial free-fall time. Star formation becomes faster and more efficient as the mass and density of the GMC increases.

Stars begin forming in the M5 and M6 clouds at $t=0.3\ t_{\rm ff}$, while in the M4 run it is delayed until $t=0.5\ t_{\rm ff}$. Because of the turbulent field, regions of the clouds have locally shorter free-fall times leading to star formation earlier than the global free-fall time. The duration of star formation is the shortest for the M4 cloud, lasting $t_{\rm sf}=0.7\ t_{\rm ff}$. The M5 cloud forms stars for a longer period in terms of its initial free-fall time, $t_{\rm sf}=1.3\ t_{\rm ff}$. M6 is still forming stars $1.3\ t_{\rm ff}$ after the onset of SF.

The SFR (Fig.~\ref{fig:global_ff}{\em (a)}) increases with cloud mass. \response{The peak SFR for the M4, M5, and M6 clouds are ${\rm SFR_{peak}}=0.4,5.5$, and~392~M$_\odot\mbox{ yr}^{-1}$ respectively. The average SFRs also increase with mass, with values of ${\rm SFR_{ave}}=0.02,\ 0.06$,~and~1.5~M$_\odot\mbox{ yr}^{-1}$.}

\response{The integrated star formation efficiency (SFE; Fig.~\ref{fig:global_ff}{\em (b)}) we discuss here is given by the ratio of stellar mass formed to the initial gas mass of the cloud $\epsilon_\star=M_\star/M_{\rm cloud}$. The instantaneous SFE is the ratio of stars to the total mass on the grid at that point in time $\epsilon_{\rm inst}=M_\star/(M_{\rm gas}+M_{\rm sink}+M_\star)$.  

For all three clouds, the instantaneous and integrated SFE closely follow each other at early times. In the M4 track, they diverge at $t\sim1 t_{\rm ff}$: $\epsilon_{\rm inst}$ increases from $35$ to $70\%$, while $\epsilon_\star$ levels out due to gas expulsion. When all gas is fully expelled from the domain all three values will converge to $\epsilon_{\rm inst}=100\%$. The instantaneous SFE for M5 is beginning to increase toward $100\%$ as the integrated SFE levels out. The $\epsilon_{\rm inst}$ for M6, however, levels out $\approx10\%$ lower than $\epsilon_\star$. This is due to the higher background density and some inflow from the boundary. Inflow is expected to occur for systems like M6, so this suggests observed SFEs of massive embedded clusters may be lower than the conversion ratio of gas to stars from the original cloud. For consistency, the best estimate for simulated SFE is the final integrated SFE value, and this is the value we use for all further comparisons to observations. We delve into the limitations of comparing observed and modelled SFEs in Sect.~\ref{sec:comparisons}.
}

The M4 cloud is representative of typical GMCs at the solar circle, and its integrated SFE lies \response{just over the top of this range at $36\%$.} Typical SFE values of nearby clusters in the Milky Way lie between 0.1--0.3 \citep{Lada2003ARA&A..41...57L}. In the higher mass clouds of M5 and M6, the SFE is much higher. The M5 cloud converted $65\%$ of its gas into stars, and the M6 cloud converted \response{89$\%$} of gas into stars. This suggests that the SFE in high-mass, high-density environments can be much higher than seen in low-mass local clusters. 

The free-fall time becomes so short in these high-mass clouds that the stellar feedback simply does not act quickly enough to stop star formation before most of the gas has formed into stars. Free-fall times of dense environments that are shorter than the development times for winds and SNe have indeed been shown to result in high SFE \citep{Dekel2023MNRAS.523.3201D}. \response{However, the more crucial factor may be that the total force budget of feedback from winds and radiation is not enough to surpass the gravity from gas and stars in dense, high-mass embedded clusters. This force-balance argument is supported by our results presented in Sect.~\ref{sec:comparisons} where the stellar feedback in a 1D model of M6 fails to expel gas from the embedded cluster's potential well. Regardless of free-fall time, in high density clouds the total feedback energy won't equal gravity until over half the cloud mass is converted to stars. However, if the free-fall time becomes so long ($t_{\rm ff}\ge10\rm\, Myr$) that SNe become a dominant form of feedback during the primary epoch of star formation, the energy from frequent SNe may start to overpower gravity and affect the SFE. We have not explored this regime, as such a high free-fall time for a $10^6\rm\, M_\odot$ cloud is rare.}

Figure~\ref{fig:global_ff}{\em (c)} shows the mass of the most massive star that has formed from random draws from the IMF. By a free-fall time, each cloud has already formed the most massive star in its cluster. We find that the mass of the most massive star increases with cluster mass. For the M5 and M6 clouds, the most massive star is at the maximum sampling mass of $100 \rm\, M_\odot$, while the M4 cloud's most massive star is \response{around $70 \rm\, M_\odot$}. This is a stochastic effect; as more stars are sampled from the IMF, you will eventually sample the most massive star in the distribution. This reproduces the effect suggested by \citet{Weidner10.1111/j.1365-2966.2009.15633.x} and \citet{Yan2023A&A...670A.151Y} that the cluster mass limits the most massive star mass. In each cloud, it is interesting to note that each instance of the formation of a very massive star, that is, above $40\rm\, M_\odot$, correlates with a slowing of star formation indicated by a reduction in the SFE slope. 

The M6 cloud forms from IMF sampling $\sim 10^6$ stars, M5 forms $\sim 10^5$ stars, and M4 forms $\sim10^4$ stars \response{, shown in Figure~\ref{fig:global_ff}{\em (d)}}. With agglomeration, the number of stars in the simulation are about $10\%$ of these numbers, so the improved version of \textsc{torch} with \textsc{petar} can simulate clusters of $>10^5$ individual stars.

\subsection{Cluster evolution}
\label{subsec:evolution}
The evolution of the global properties of the formed star clusters occurs quite similarly for all three clouds, but the magnitude of their values depends greatly on the cloud's initial mass.

\response{The stellar velocity dispersion (Fig.~\ref{fig:global_ff}{\em (e)}) generally increases with initial cloud mass.} The velocities of stars increase at a slow pace before leveling out after $1\,t_{\rm ff}$. For the M4 cluster, the velocity dispersion levels out at just \response{1.0~km~s$^{-1}$. At late times, the velocities of the stars begin to slightly decrease in M4. This decrease correlates to the increasing half-mass radius of M4, indicating the star cluster is expanding and the stellar velocities are slowing.} The M5 cluster reaches a velocity dispersion of 5~km~s$^{-1}$ and the M6 cluster has a velocity dispersion of 20~km~s$^{-1}$. 
The deeper potential wells of the higher mass clusters, going as the square root of the mass for these similar sized objects, drive the higher velocity dispersions\response{, although the measured dispersion increases somewhat faster with mass than the potential well depth}. In the case of M6, the potential well depth exceeds the sound speed of ionized gas, preventing gas from escaping even after ionization. \response{For M6, the average sound speed in cells where the gas is fully ionized is $c_{\rm ion}= 16.4\rm\, km~s^{-1}$.}

The evolution of the half-mass radius $R_{1/2}$ of all the stars in the cluster (Fig.~\ref{fig:global_ff}{\em (f)}) seems to be split into a high and low-mass regime. The M5 and M6 clusters follow the same track closely. \response{From $0.3\,t_{\rm ff}$ to $1.0\, t_{\rm ff}$, $R_{1/2}$ increases to a peak of $\sim3.75$\,pc at $\sim 0.6\,t_{\rm ff}$ then goes down to $\sim1$\,pc (M5) and $\sim0.5$\,pc (M6) and begins to level out.} The similarities in the evolution of the two clusters are most likely due to the fact that both clouds at early times form enough stars for distinct sub-clusters to form and merge. The sub-clusters have formed in the same places so both clusters peak at roughly the same $R_{1/2}$. \response{All of the stars in M5 and M6 remain bound, suggesting the clusters are relaxing into gravitational equilibrium.} Longer runs following just the stars after gas dispersal will ultimately be needed to demonstrate this. \response{M6 is a smaller and denser cluster than M5, likely due to the much deeper potential well of M6. Similar stellar densities are observed in the super-star clusters (SSCs) of starburst NGC 253 \citep{Rico-Villas2020MNRAS.491.4573R} with sizes of $R<1.7\rm\, pc$ and stellar masses of $10^{4-6} M_{\odot}$.

The M4 cluster grows slightly differently. It increases to a peak of $R_{1/2}=2~\rm pc$ at 1 $t_{\rm ff}$, decreases to $R_{1/2}=1\rm~pc$, then linearly increases to 2.5~pc by 1.3~$t_{\rm ff}$. The M4 cluster is \response{expanding}, but $85\%$ of its stars remain bound, so complete dissolution has not yet occurred (Fig.~\ref{fig:global_ff}{\em (i)}).}

The onset and duration of gas dispersal from the star clusters depends strongly on the initial mass and density of the cloud. Figure~\ref{fig:global_ff}{\em (g--i)} shows the time evolution of the mass, virial parameter, and bound mass fraction of the gas and stars. With these three plots we can track the degree of gas dispersal. From the mass plot we see that by the end of star formation, \response{ the M4 and M5 clusters have expelled a significant fraction of the initial cloud. Only $10\%$ of the original gas mass remains in M4 and $<10\%$ in M5. Both M4 and M5 are well on their way to full gas expulsion, as in both clusters $<10\%$ of the gas still on the grid is bound.} The gas in the M4 and M5 systems become super-virial by a free-fall time. The gas in M5 takes $\sim 10\%$ longer to become unbound, but progresses identically to the M4 gas. The gas in M6 differs significantly as it remains sub-virial even beyond one free-fall time. The potential well created by the massive cluster is enough to keep the remaining gas infalling\response{, suggesting star formation is not yet quenched in the M6 cluster.}

The only star cluster that is starting to disperse is the M4 cluster (Fig.~\ref{fig:global_ff}{\em (i)}). The other two star clusters remain fully bound. \response{The stars in M5 and M6 remain sub-virial, while the stars in M4 just reach virial equilibrium by the final simulation time (Fig.~\ref{fig:global_ff}{\em (h)}).} The dispersal time of the gas and stars increases with initial cloud mass as there is more gravity for the stellar feedback to counteract. Although massive clusters have more stars injecting feedback, the increasing gravity overpowers the feedback. At high densities, where the potential well depth exceeds the sound speed of ionized gas, ionization feedback cannot disperse gas, while the short free-fall time assembles dense gas more quickly than feedback can work against gravity.

\begin{table}
\caption{Results.}   
\label{table:results}   
\centering   
\begin{tabular}{l|ccccc}          
\hline\hline                        
Run & M$_\star$ & $N_\star$ & \response{$\epsilon_\star$} & $\rm \langle SFR \rangle$ & $\rm SFR_{pk}$ \\   
\hline                                   
    M4 & \response{3,628} & \response{6,488} & \response{0.36} & \response{0.022} & \response{0.4}\\ 
    M5 & 64,733 & 112,661 & 0.65 & 0.063 & 5.535\\ 
    M6 & 845,815 & 1,468,969 & 0.85 & \response{1.530} & 392.0\\ 
\hline  
\end{tabular}
\tablefoot{Column values and units: Stellar mass $[\rm\, M_\odot]$, number of stars formed from IMF sampling (number of stars in the simulation after agglomeration of stars $<4\rm\, M_\odot$ is $\sim10\%$ of this value), \response{integrated} SFE $[M_\star/M_{\rm cloud}]$, average SFR $\rm\, [M_\odot\ yr^{-1}]$, peak SFR $\rm\, [M_\odot\ yr^{-1}]$.}
\end{table}

\section{Discussion} \label{sec:discussion}

\subsection{Limitations of comparing observed and modelled star formation efficiencies}
\label{sec:comparisons}

\response{

We compare the SFEs of our modelled clusters to observations by using the integrated quantity $\epsilon_\star$, the stellar mass divided by the initial cloud mass. This is the total fraction of cloud mass that has been converted into stars. However, this value is impossible to calculate for observed clusters, as the only information available is how much gas and stars are present within a certain area. Thus, observations of star clusters only quote the instantaneous value $\epsilon_{\rm inst}$. There is a 3--4 order of magnitude spread in the observed SFE and SFR of Galactic GMCs \citep{Lee2016ApJ...833..229L}.

A proper comparison between simulated and observed SFEs requires accounting for the amount of ongoing star formation, and determining whether the embedding gas is collapsing or dispersing. Comparisons done without accounting for the evolutionary stage of the cluster are misleading. 
The apparent SFE ranges from 0--100\% over the lifetime of every star cluster that reaches a gas-free state, which may explain the spread in observed SFE. This issue is starting to be explored. \citet{Geen2017MNRAS.471.4844G} suggest techniques for converting observed to integrated SFEs, although conclude that this is non-trivial and find errors up to a factor of 10. They find overall higher observed SFEs than integrated SFEs when they applied observational techniques to their simulations. On the other hand, \citet{Grudic2019MNRAS.488.1501G} find lower observed than integrated SFEs in their models due to the inaccuracies of techniques for estimating stellar mass. One example of this comes from using only the young stellar population as a tracer for stellar mass, which underestimates the total stellar mass. They also discuss variability in observed SFEs due to the changing SFE over the course of a GMC lifetime, from first star formation to gas dispersal.

Further studies must be dedicated to outlining a systematic way to convert between observed SFEs and the final integrated SFEs we define in our models. Until then, direct comparisons should be interpreted with caution. 
}

\subsection{Observations}

Galactic surveys of embedded clusters in the Milky Way typically find the SFE to be $\lesssim 30\%$ \citep{Lada2003ARA&A..41...57L}, with some studies finding lower values of $\lesssim 8\%$ \citep{Evans2009ApJS..181..321E,Peters2011ApJ...729...72P}. The M4 cloud, which is a good representative of galactic GMCs\footnote{See \citet{Rice2016ApJ...822...52R} for a catalog of Milky-Way molecular cloud properties.}, agrees well with this SFE albeit at the high end of observed values. This could be due to the low virial parameter used, which is appropriate for M6 but lower than the average \response{observed} value of $\alpha_v=1$ seen in \response{clouds similar to} M4. \response{The missing radiative feedback from stars $<20~\rm M_\odot$ could also be a factor causing the high SFE of M4 given the low-density of the cloud. Feedback contributions from low-mass stars could be a key factor in quenching star formation in Milky Way-like clouds.}

The higher mass \response{clouds similar to M5 and M6}, however, have SFEs well above $30\%$. While there are no Milky-Way analogs to the M6 cloud, there are a few for M5. There is the W43 GMC with $1.32\times 10^5 \rm\, M_\odot$ of gas within $R\sim10\rm\, pc$ \citep{Lin2016ApJ...828...32L}, similar to the M5 cloud with $R=11.7\rm\, pc$. The W49 star forming region has a central YMC with stellar mass $\gtrsim 5\times 10^4\rm M_\odot$ and gas mass $\sim 2 \times 10^5\rm M_\odot$ and $\sim 1.1 \times 10^6\rm M_\odot$ within 6 and 60~pc respectively \citep{GalvanMadrid2013ApJ...779..121G}. This gives a current SFE of $20\%$ in the inner region. \response{W49 has ongoing star formation, and its embedded gas cloud is twice as massive and 15 times as dense as our M5 model. Based on our results for M5 and the fact that SFE increases with density, we predict W49 will exceed the SFE found for M5 of $\epsilon_\star = 65\%$.}

Though conditions required to form the M6 cloud are not observed in the Milky Way, they are present in other galaxies. \response{Starburst galaxies have been observed to host SSCs with SFEs of $\epsilon_{\rm inst}\approx50-80\%$. These SSCs cover a size and mass range comparable to our models (see Table~\ref{table:obs}).}

\begin{table}
% \tablewidth{\}
\caption{Properties of observed super star clusters.}   
\label{table:obs}   
\centering   
\begin{tabular}{l|ccc}          
\hline\hline                        
galaxy & NGC 5253 & NGC 253 & NGC 4945 \\
\hline 
    type & dwarf & starburst & starburst \\
    $M_\star\rm\, [M_\odot]$ & $1.1^{+0.7}_{-0.2}\times10^6$ & $10^{4-6}$ & $10^{4.6-5.7}$ \\
    $R\,\rm [pc]$ & $28\times 52$ & $0.34-1.7$ & $1.4-4.0$ \\
    $\epsilon_{\rm inst}$ & $61^{+84}_{-16}\%$ & $30-90\%$ & $>50\%$ \\
    $N_{\rm SSC}$ & 1 & 12 & 27 \\    
\hline  
\end{tabular}
\tablefoot{\response{Characteristics of super star clusters in the galaxies} NGC 5253 \citep{Turner2015Natur.519..331T}, NGC 253 \citep{Leroy2018ApJ...869..126L,Rico-Villas2020MNRAS.491.4573R}, and NGC 4945 \citep{Emig2020ApJ...903...50E}.}
\end{table}

% The SSC masses in NGC 253 hosting starbursts include NGC 253 \citep{Rico-Villas2020MNRAS.491.4573R}, 

Additionally, the disks of gas-rich high redshift galaxies can be violently unstable and are thought to form clouds similar to M6 \citep[see][]{Tacconi2020ARA&A..58..157T}. We can now directly observe the high redshift environment of forming GCs with JWST. \response{\citep{Li2023arXiv231114662L} predicts JWST will find feedback-free starbursts \citep%[FFBs;][]
{Dekel2023MNRAS.523.3201D}, which are massive galaxies at $z\gtrsim10$ with high SFEs due to dense gas with free-fall times $\leq10\rm\, Myr$ forming stars effectively free of stellar feedback.} Recent JWST observations uncovered ``younger'' populations of GCs in galaxies at redshift $z=0.38$ \citep{Harris2023MNRAS.526.2696H}, and \response{others} are expected to observe GCs up to $z=1$ without lensing \citep{ReinaCampos2024MNRAS.531.4099R}. 

\response{
With lensing, clumps that are likely proto-GCs can be observed at redshift $z > 1$ \citep{Adamo2023MNRAS.525L...6A,Claeyssens2023MNRAS.520.2180C}. One such proto-GC candidate was found through lensing at $z\sim6$ with $\lesssim10^6\rm\, M_\odot$ and a core radius of $R_c<13\rm\, pc$ \citep{Vanzella2019MNRAS.483.3618V}. Another more massive bound YMC $3\rm\, Myr$ old was found at $z=2.37$ with $\sim10^7\rm\, M_\odot$, and $R\sim 8\rm\, pc$ \citep{Vanzella2022A&A...659A...2V}. A strongly lensed galaxy at $z=4$ contains three bound YMCs each younger than $<30\rm\, Myr$ with masses between (0.7--4.0)$\times 10^6 \rm\, M_\odot$ and radius estimates of 3--20~pc \citep{Vanzella2022ApJ...940L..53V}. The Sunrise arc is a strongly lensed $z\approx6$ galaxy found to contain six YMCs with masses $\sim10^{6-7}\rm\, M_\odot$, radii of $\sim$1--20~pc, and ages 1--30~Myr \citep{Vanzella2023ApJ...945...53V}. Most of these recently discovered YMCs or proto-GCs are analogous in size and mass to the M6 cluster or larger. Dense regions of prolific star formation that form these objects seem pervasive in the early Universe, and more will surely be discovered as more JWST data arrives. Due to their similar properties, we argue that these clusters formed in the same manner as M6. 
}

Another situation that can form GMCs \response{similar to M6} is major galaxy mergers with small mass ratios. Tidal interactions of major galaxies are linked to bursts in star formation \citep{Larson1978ApJ...219...46L,Lonsdale1984ApJ...287...95L,Barton2000ApJ...530..660B,Ellison2008AJ....135.1877E,Renaud2019A&A...625A..65R}. Since most massive galaxies are believed to undergo at least one merger in their lifetime, this is not a rare occurrence. Galaxy mergers have been suggested as the progenitors of YMCs and younger GCs \citep{ashman1992,vandenBergh2001ApJ...559L.113V}. We note that this only applies to major galaxy interactions: minor galaxy interactions with large mass ratios produce little to no enhancement of the overall SFR \citep{Cox2008MNRAS.384..386C,Tress2020MNRAS.492.2973T}.

In the interacting Antennae galaxies, the Firecracker cloud, which resembles M6, was observed by \citet{Whitmore2014ApJ...795..156W}. \citet{Finn2019ApJ...874..120F} constrains its mass and characteristic radius to (1--9)$\times 10^6\rm\, M_\odot$ and $22\ \rm\, pc$. The Firecracker cloud is in the very early stages of star formation, as it is estimated to have only formed $M_\star\lesssim 10^4\rm\ M_\odot$. This is less than $10\%$ of the expected stellar mass of the final star cluster \citep{Johnson2015ApJ...806...35J}. These observations show that progenitor clouds similar to M6 can form before any significant amount of star formation occurs. 

A survey of the molecular clouds in the Antennae galaxies done by \citet{Wei2012ApJ...750..136W} revealed two populations of MCs, with a distinct break in the differential mass function at $\log(M_{\rm cloud}/\rm M_\odot)=6.5$. Clouds above this mass were found in the regions of intense star formation, while the lower mass clouds were in more dormant regions. The large velocities seen in the high SF regions suggest compression by shocks, supporting the idea that galaxy mergers lead to high-mass GMCs that become sites of extreme star formation.

\begin{figure}
\centering
    \includegraphics[width=\hsize]{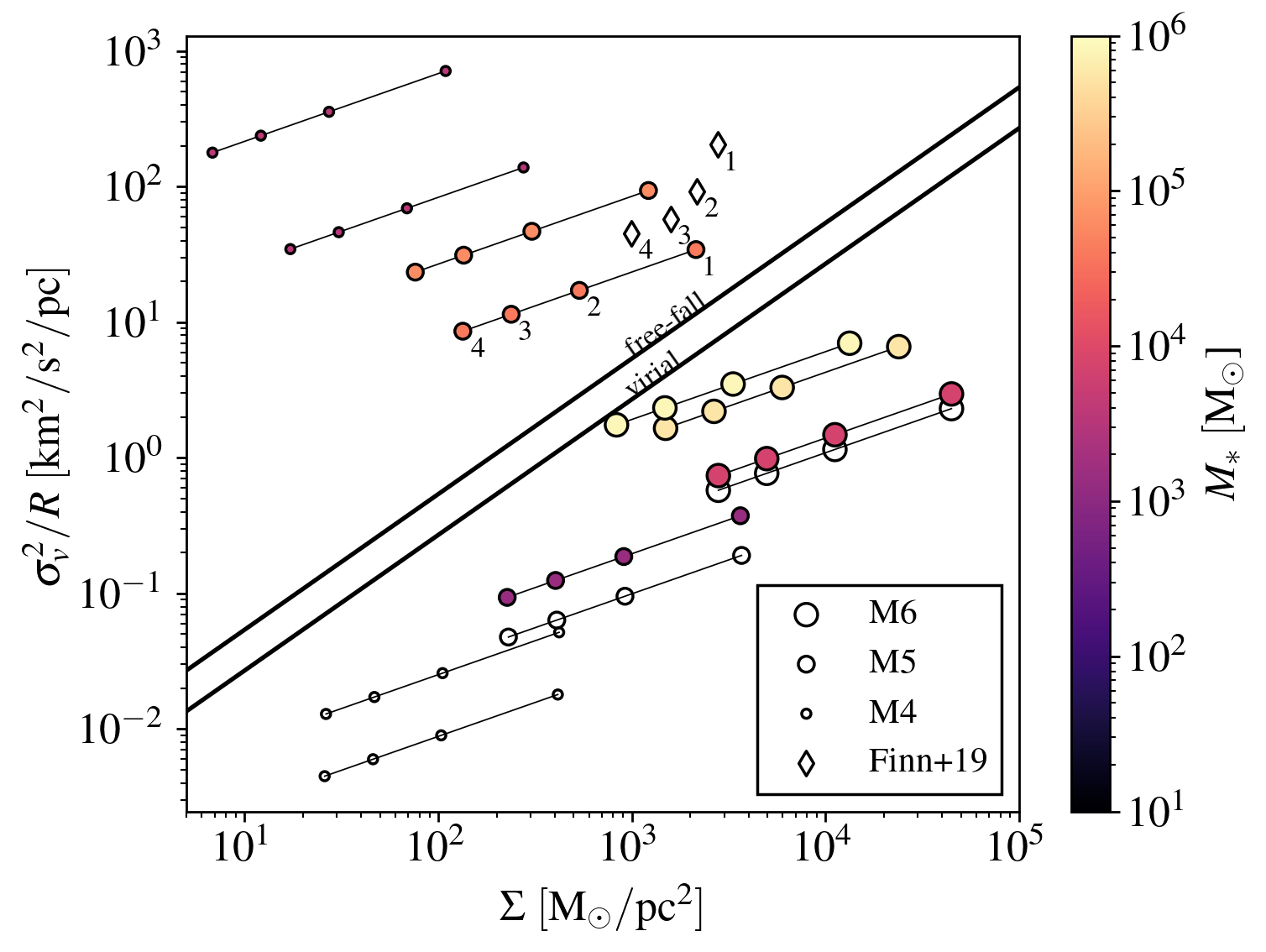}
    \caption{Size-line width coefficient versus surface density for the Firecracker cloud \citep{Finn2019ApJ...874..120F} and the M4, M5, and M6 clouds. The four red diamonds are observations, done using apertures (1--4) with radii $R =$ 6.4, 15, 26, and 37 pc. The other points are our simulations, with connected points corresponding to apertures (1--4) of radii $R =$ 5, 10, 15 and 20 pc. The colors indicate how much stellar mass has been produced (with empty points indicating no star formation). The lines correspond to virial equilibrium and free-fall as labeled \response{\citep[see Fig.~2 of][]{ballesteros2011}}.}
    \label{fig:firecracker}
\end{figure}

\citet{Finn2019ApJ...874..120F} measured the velocity dispersion in the Firecracker cloud and found it to be neither in virial equilibrium nor free-fall. They conclude that there must be a high pressure background to contain the gas at such high densities in equilibrium. 

We compare this velocity dispersion to those in our clouds over time to test whether we reach such high velocities through the addition of stellar feedback to free-fall collapse alone or whether a high pressure background is indeed needed. The results of this comparison are shown in Figure~\ref{fig:firecracker}. We plot the size–line width coefficient $\sigma_v^2/R$ against surface density $\Sigma=M/\pi R^2$ for each of our clouds to compare with the observations from \citet{Finn2019ApJ...874..120F}. The observations are shown by the red points in Figure~\ref{fig:firecracker} corresponding to 4 aperture sizes they used for $R$: 6.4, 15, 26, and 37 pc. We use four smaller aperture sizes: 5, 10, 15, and 20 pc, as our cloud is half the radius of the Firecracker. We plot four times for each of our simulations corresponding to $t=0.0,0.5,1.0,1.25\, t_{\rm ff}$, with a line connecting the points showing the set of apertures. The apertures for each set increase from right to left with decreasing surface density. The colors of the points indicate the amount of stellar mass formed, colored white when no SF has occurred yet. The sizes indicate the initial gas mass of the cloud, with the smallest being the M4 simulations and the largest being M6. \response{The two labelled lines on the plot indicate the analytical conditions for free-fall and virial equilibrium \citep[see the discussion of Fig.~2 in][]{ballesteros2011}. We include these lines for easier comparison to the observations presented by \citep{Finn2019ApJ...874..120F} rather than for inferring the state of the system. Unintuitively, velocity dispersions in a state of free-fall exceed those of virial equilibrium. }

Though the Firecracker cloud is as massive as M6, its surface density is \response{lower}: more comparable to M5. \response{This is why we see an overlap between M5 and the Firecracker cloud when M5 has formed $4\times 10^4\rm\, M_\odot$ of stars, which is on the order of the stellar mass estimated to have already formed in the Firecracker cloud of $\leq 10^4\rm\, M_\odot$. This aligns with the idea that surface density is more influential than mass in the formation of star clusters.} This also supports the possibility that the Firecracker gas velocities could be caused by the contribution of stellar feedback to the velocity dispersion in addition to free-fall collapse. The fast dynamical evolution of our models suggests that these are not equilibrium objects, making it unnecessary to invoke a high-pressure background to keep the cloud from expanding. This suggests objects like the Firecracker cloud can form from collapse with observed velocity dispersions without invoking a high pressure background medium. 

\subsection{Other simulations}
\label{sec:compare_obs}

For massive star clusters to form, they must survive the epoch of gas dispersal and remain bound. \response{An analytical model predicts a positive correlation of SFE to SFR and initial cloud mass \citep{Zamora-Aviles2014ApJ...793...84Z}. At \response{high enough surface} densities, stellar feedback cannot compete with star formation.} Numerical studies done by \citet{Geyer2001MNRAS.323..988G} found that if the stars are initially in virial equilibrium with the remaining gas, only clusters with SFE $\ge 50\%$ remain bound against the outflow of the gas. \citet{Li2018ApJ...861..107L} finds in their cosmological galaxy formation models that though galactic properties are unaffected by varying $\epsilon_\star$, the properties of star clusters are \response{affected. In particular}, they found that the initial bound fraction of stars increases with $\epsilon_\star$ and cloud mass. 
\citet{Farias2023MNRAS.tmp.3530F} ran cluster formation models from $2\times 10^4\rm\, M_\odot$ clouds and finds that SFE and gas expulsion time correlate with global bound fraction, with all SFEs $\leq 20\%$ and all bound fractions $\leq 40\%$.
Although it is still possible to form bound clusters with low SFE, these studies imply massive bound star clusters were most likely formed with high SFEs. 

\citet{Menon2023MNRAS.521.5160M} also finds high SFEs of $\sim80\%$ for $10^6\rm\, M_\odot$ clouds with feedback in the form of radiation pressure solved using a variable Eddington tensor approach as opposed to our ray-tracing method. In this density regime, radiation pressure is the dominant feedback mechanism\footnote{See extended data Figure 5 of \citet{Howard2018NatAs...2..725H} and Figure 12 of \citet{Krumholz2019ARA&A..57..227K}.}. They conclude that radiation pressure simply cannot regulate star formation for clouds with surface densities $\Sigma \gtrsim 10^3\rm\, M_\odot\ pc^{-2}$. 

Our results are more constraining: We included more feedback physics, and we still achieved SFEs of $\epsilon_\star>80\%$. Our M6 cloud is above this surface density with $\Sigma=2.3\times 10^3 \rm\, M_\odot\ pc^{-2}$. They also tested a larger $10^6\rm\, M_\odot$ cloud with roughly the same surface density as our M5 cloud, and find an SFE of $\epsilon_\star\approx60\%$ comparable to the SFE of our M5 cluster of $\epsilon_\star=65\%$.

Other simulations of massive star cluster formation with initial cloud mass of $10^6\rm\, M_\odot$ find high SFEs of $\sim 65\%$ \citep{Grudic2018MNRAS.475.3511G} and $38\%$ \citep{Kim2018ApJ...859...68K} for surface densities of $\Sigma=1.27 \times 10^4 \rm\, M_\odot\, pc^{-2}$ and $\Sigma=500\rm\, M_\odot\, pc^{-2}$, respectively. \citet{Kim2018ApJ...859...68K} finds an even higher SFE of $51\%$ for a $10^5\rm\, M_\odot$ cloud but with a surface density of $\Sigma=1.27 \times 10^3\rm\, M_\odot\, pc^{-2}$. \response{Cluster models in \citet{Kimm2022ApJS..259...21K} reached $\epsilon_\star=$50--72\% from an initial cloud of $1.4\times10^6\rm\, M_\odot$ and $\Sigma=647\rm\, M_\odot\, pc^{-2}$ despite SN and radiation feedback. Another cluster, modeled with radiation feedback by \citet{Fukushima2021MNRAS.506.5512F}, reached $\epsilon_\star=78\%$ from a $10^6\rm\, M_\odot$ cloud with $\Sigma=3.2 \times 10^3 \rm\, M_\odot\, pc^{-2}$. This study also finds that bound cluster formation only occurs with $\epsilon_\star\geq30\%$. 
\edit{Recent models of $10^6\rm~M_\odot$ clouds with $R=10\rm~pc$, $\alpha_v=0.1$, and $Z=0.2~Z_\odot$ reached $\epsilon_\star=50\%$ \citep{Fujii2024arXiv240606772F}.}
The SFE in all of these studies increases strongly with surface density and slightly with initial cloud mass. Our results combined with those from previous models provide evidence that the formation of bound YMCs requires not only a high cloud mass but also, and more importantly, a high surface density. 
}

\response{Simulations of star clusters forming from clouds similar to our M5 cloud resulted in lower SFEs of 10--30\%. The main difference between these models and ours is the use of sink particles to represent sub-clusters with combined feedback compared to our tracking of feedback from individual massive stars. A cloud modelled by \citet{He2019MNRAS.489.1880H} with an initial mass of $10^5\rm\, M_\odot$, peak number density $n=1.8\times 10^3\rm\, cm^{-3}$, metallicity $\rm Z=Z_\odot$, a higher virial parameter $\alpha_v=0.4$, and stellar feedback only through UV radiation reached a SFE of $13.7\%$ by $6\, t_{\rm ff}$. The cloud in \citet{Ali2021MNRAS.501.4136A} is the same mass and metallicity as M5, almost the same radius, $R_{\rm cloud}=11.9\rm\, pc$, but is initially super-virial, with $\alpha_v=1$. By $0.75\, t_{\rm ff}$, the SFE reached only $10\%$ while M5 reached $\epsilon_\star=20\%$ by this time. This difference may come from the different initial virial parameters, which prolongs the formation of the cluster in \cite{Ali2021MNRAS.501.4136A}. Another possible cause is different feedback models. Injecting stellar feedback from entire sub-clusters rather than individual stars could artificially strengthen the effect of feedback resulting in a lower SFE.
}

\edit{\citet{Fujii2021PASJ...73.1074F} presents star-by-star cluster models with feedback in the form of radiation, radiation pressure and stellar outflows. For an initial cloud of $10^5\rm~M_\odot$, $R=20~\rm pc$, and $\alpha_v=0.25$ they find $\epsilon_\star=40\%$. The same cloud with a larger virial parameter of $\alpha_v=1.0$ only reached $\epsilon_\star=40\%$. Their sub-virial model agrees with our findings for M5. The super-virial model with a lower SFE further indicates that the high SFEs we find are possibly due to the low initial virial parameter, particularly for lower density clouds.}

\response{A colliding flow model of star formation in GMC environments described in \cite{Colin2013MNRAS.435.1701C} with individual star formation and ionizing radiation found SFEs of $\epsilon_\star=10$--30\% depending on the degree of concentration by the flows. The two cylindrical streams were very large, with $r=64 \mbox{ pc and }\ell=112\rm\, pc$ and rarefied, with $n=1\rm\, cm^{-3}$, with the total mass in the two streams equalling $9\times10^5\rm\, M_\odot$. The different initial conditions hinder a direct comparison to our SFE values, but reaching high SFEs from low density flows aligns with our results for M4.}

\response{Simulations have broadly found star formation to be suppressed with each additional form of stellar feedback included in the model \citep[see ][]{Dale2015NewAR..68....1D}. The exclusion of protostellar jets in our feedback model may artificially raise the SFR, as they contribute to the dispersal of gas around even low-mass stars at small scales \citep{Chevance2023ASPC..534....1C}.} Due to the quantity of low-mass stars and the collimated shape of the outflow, jets are drivers of turbulence at large scales in GMCs \citep[e.g.,][]{Nakamura2007ApJ...662..395N,Fedderath2015MNRAS.450.4035F,Appel2022ApJ...927...75A}. 
\response{These models do} show that jets are an important factor in slowing the growth rate of the integrated SFE, though the final SFE is not known due to the duration of the simulations \citep{Fedderath2015MNRAS.450.4035F}.
\citet{Guszejnov2021MNRAS.502.3646G} performed simulations of star-by-star cluster formation from  $2\times 10^4\rm\, M_\odot$ clouds with stellar feedback, including protostellar jets as well as radiation, winds, and SNe. Simulations were repeated that isolated each form of feedback. They found jets to be important in regulating the growth of low-mass stars and constraining the IMF. Radiation and jets were the primary form of feedback that slowed star formation and dispersed the cloud. However, \response{again} the simulations were not run until the end of star formation, so the degree to which each \response{effect changes} the final SFE remains uncertain.

Despite their ubiquity, \response{jets} cannot prevent gas in high-density GMCs from forming stars eventually nor contain the power needed to disperse GMCs \citep[see][]{Chevance2023ASPC..534....1C}. 
%[moved] On the other hand, studies also show that jets are an important factor in slowing the growth rate of the integrated SFE, though the final SFE is not known due to the duration of the simulations \citep{Fedderath2015MNRAS.450.4035F,Appel2022ApJ...927...75A}. \response{However, most other simulations of clouds similar to M4, like those highlighted in \cite{Chevance2023ASPC..534....1C}, are initialized to be super-virial or in virial equilibrium, while our model begins sub-virial. Either or both of the sub-virial initial state and the lack of protostellar jets could be the reason for the higher SFE in M4 compared to other models of similar mass and surface density clouds.} 
% [moved up]\citet{Guszejnov2021MNRAS.502.3646G} performed simulations of star-by-star cluster formation from   $2\times 10^4\rm\, M_\odot$ clouds with stellar feedback, including protostellar jets as well as radiation, winds, and SNe. Simulations were repeated that isolated each form of feedback. They found jets to be important in regulating the growth of low-mass stars and constraining the IMF. Radiation and jets were the primary form of feedback that slowed star formation and dispersed the cloud. However, the simulations were not run until the end of star formation, so the degree to which each affect the final SFE is uncertain. 
% [eliminate redundancy] This suggests the SFE of M4 may be overestimated due the exclusion of jets. 
The effect that jets would have on more massive clouds remains unclear. A\response{lthough they may indeed slow star formation to observed values, as proposed by \citet{Chevance2023ASPC..534....1C}, a}nalytic work by \citet{matzner2002} suggests that more massive clouds would be resistant to dispersal by jets, consistent with simulations by \citet{guszejnov2022}.

\begin{figure}
 \centering
     \includegraphics[width=\columnwidth]{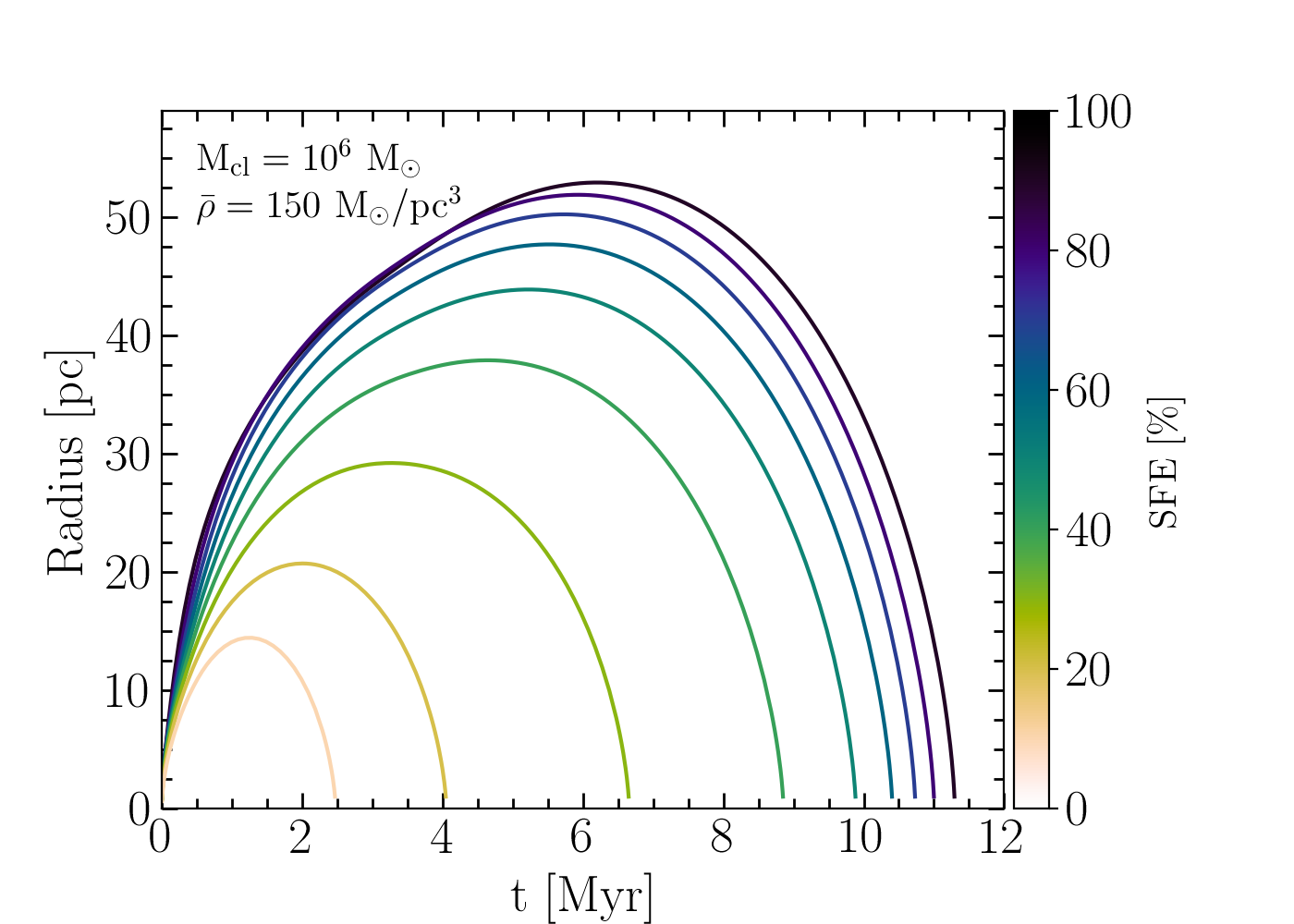}
     \caption{\textsc{warpfield} evolution of shell radius versus time with different initial SFEs with the same parameters described in M6. In all cases, stellar feedback is inefficient in dispersing the surrounding dense cloud, and the shell eventually undergoes re-collapse.}
     \label{fig:warpfield}
 \end{figure}

\response{In order to verify the physical plausibility of the high SFE in M6 despite the large number of formed stars}%After the M6 simulation resulted in such a high SFE, we wanted to verify that this many stars would not just immediately blow away the gas. To do this
, we have directly compared our 3D results to a followup calculation using the one-dimensional (1D) code Winds And Radiation Pressure: Feedback Induced Expansion, colLapse and Dissolution  \citep[\textsc{warpfield};][]{Rahner2019MNRAS.483.2547R}. This code models the effect of stellar feedback from young clusters on their natal gas cloud \response{in spherical symmetry}. \textsc{warpfield} is designed to solve for the self-consistent motion of a 1D spherical gas shell evolving under the influence of feedback mechanisms including stellar winds, SNe, and radiation pressure, with consideration of gravity. We ran \textsc{warpfield} using the same \response{initial} conditions as chosen for the M6 run (i.e., mass, density, temperature), with the addition that we varied the SFE from $\epsilon_\star = 0.1$--0.9 in bins of 0.1, as shown in Figure~\ref{fig:warpfield}. \response{This varies the strength of the stellar feedback to test whether the M6 cloud would still be stable given amount of stars formed.

For all SFE values, the shocked gas eventually re-collapses. At this high density, the included feedback is not strong enough to completely disperse the cloud. The SFE of our M6 cluster most closely resembles the \textsc{warpfield} runs with $\epsilon_\star=80\%\ \&\ 90\%$ in Figure~\ref{fig:warpfield}. The \textsc{warpfield} model reaches a maximum radius of $R\sim55\rm\ pc $ at $t\sim6.5\rm\ Myr$, and collapses back to $R=0\rm\ pc$ by $t\sim11.25\rm\ Myr$. In the 1D model, all feedback occurs at a single point, so it is more effective than in our star-by-star 3D model, as in multiple dimensions channels that vent thermal energy can exist. Nevertheless, the gas still re-collapses, promoting further star formation. The expanding gas is not accelerated fast enough to escape the deep potential well of the massive cloud and the cluster that forms from it. This result supports the idea that the high SFE is due to the total feedback strength being weaker than gravity at these densities.
}

However, our results do suggest that more dispersed star formation leading to increased energy dissipation by radiative cooling may not even allow that much expansion. To resolve SFE well, feedback must be modelled for individual stars instead of for entire clusters. Approximating feedback as a sum for an entire cluster underestimates the SFE.

\section{Conclusions} \label{sec:conclusions}

We performed numerical simulations of star cluster formation from gas clouds that run until star formation ceases or slows significantly due to stellar feedback dispersing any remaining gas. We tested initial cloud masses of $10^4,\ 10^5,\ \mbox{and}\ 10^6\rm\, M_\odot$ with radius $R=11.7\rm\, pc$, holding all other characteristics of the initial cloud and simulation parameters the same. We analyzed the star formation histories and followed the evolution of the gas and forming star clusters. From this study, we conclude the following:

\begin{itemize}
    \item[--] Giant molecular clouds with surface density $\Sigma \ge 10^2 \rm\, M_\odot\ pc^{-2}$ and mass $M_{\rm cloud} \ge 10^5\rm\, M_\odot$ can form fully bound star clusters with stellar mass $M_\star \ge 10^4 \rm\, M_\odot$ with a high SFE $\epsilon_\star \ge 65\%$ over a short time $t_{\rm sf} \approx 1 t_{\rm ff}$, as seen by M5 and M6. The lower mass and density M4 cloud forms a cluster with a lower bound mass fraction of $60\%$.
    \item[--] The Firecracker cloud in the Antennae galaxies, with a mass of 1--9$\times 10^6\rm\, M_\odot$ and a radius of $22\rm\, pc$ \citep{Finn2019ApJ...874..120F}, is a close analog to our M6 cloud, though with a surface density more closely matching our M5 cloud. From our results we can estimate that the Firecracker cloud will convert 65--85\% of its mass into stars within a free-fall time and that it will form a YMC.
    \item[--] It has been suggested that the Firecracker cloud must be surrounded by a high pressure medium to contain it because of its high surface density and size-line width coefficient $\sigma_v^2/R$ \citep{Johnson2015ApJ...806...35J,Finn2019ApJ...874..120F}. However, the M5 cluster reaches the same values by the time it forms $M_\star\approx10^4\rm\, M_\odot$ worth of stars, the same amount of stellar mass estimated to have formed in the Firecracker cloud. This suggests another possibility: Rather than being an equilibrium object confined in a high pressure environment, the Firecracker cloud is actually dynamically collapsing and forming stars, and the high velocity dispersion of the gas is from the combination of free-fall collapse and stellar feedback. 
    \item[--] Star formation from GMCs is capable of achieving up to $85\%$ efficiency at high densities. Our M6 cloud is the most efficient of our models, converting $\epsilon_\star=85\%$ of its gas into stars. Even with hundreds of massive stars producing feedback, the short timescale of gravitational collapse for dense massive clouds renders the stellar feedback inefficient at slowing early star formation. However, even at much lower densities and masses, the M5 and M4 clouds achieved high SFEs of $\epsilon_\star=65\%$ and \response{$36\%$, respectively. In dense, massive clouds, the total dispersing force of stellar feedback from winds and radiation cannot counteract the gravity from stars and gas until over half the cloud mass is converted into stars.}
    \item[--] The M4 cloud has a typical mass and size of Milky-Way GMCs. The SFE of M4 matches the maximum observed SFE values. This high SFE could be because of the low initial virial parameter of the cloud\response{, or it could be due to the missing FUV radiation from stars $<20~\rm M_\odot$.} Alternatively, the exclusion of the protostellar jet feedback mechanism may be important for clouds similar to M4 clouds, as suggested, for example, by \citet{Chevance2023ASPC..534....1C}. Further studies must be done to constrain the effect of varying the virial parameter and including protostellar jets on integrated SFE.
    \item[--] Star formation is fast in our models of clouds with low $\alpha_v$. Regardless of the initial mass or density, the majority of star formation occurs within the first global free-fall time of the collapsing GMC. 
    \response{Collapse occurs and stars are produced so rapidly that stellar feedback is only prevalent and strong enough to clear dense gas from the cluster's deep potential well after most of the cloud has formed into stars.} 
    The speed of star formation may depend strongly on the initial virial parameter and the inclusion of jets.
    \item[--] A 1D stellar feedback model \textsc{warpfield} was run using the same mass and density as the M6 simulation. In it, the gas re-collapses even for SFEs up to $90\%$. Even centralized feedback cannot expel the gas from the potential well of the massive cluster that forms. The \textsc{warpfield} results indicate that the expanding gas shell for $\epsilon_\star=85\%$ collapses back to $R=0$ by 11 Myr. 
    \item[--] Including feedback for individual stars rather than adding the total energy for the cluster at a single point is important for correctly constraining star formation histories. Modelling individual stellar feedback spreads the feedback energy enough to greatly reduce its effectiveness at clearing the natal gas because of the resulting enhanced radiative cooling. Models that add stellar feedback for the entire star cluster at a single point appear to overestimate the effect of the feedback on the gas and the star formation timescale and to underestimate the final SFE.
\end{itemize}

In conclusion, bound massive star clusters such as YMCs and GCs readily form from high-mass, dense GMCs. The GMCs can become this dense and massive naturally, even in the present day, as shown by the Firecracker cloud in the Antennae galaxies. In the early Universe, where galaxies were much more gravitationally unstable, these conditions would be much more common. The subsequent star formation from these dense high-mass clouds is highly efficient, converting $\ge 40\%$ of the gas mass into stars \response{within the first free-fall time of the initial cloud}. The short timescales of star formation \response{and/or the deep gravitational potential wells of dense, massive clouds} render stellar feedback unable to significantly slow star formation, leading to integrated efficiencies as high as $85\%$ for more massive clouds. After their formation, the clusters born in these environments remain bound after $90\%$ of the gas is expelled. 

\response{Until recently, directly observing proto-GCs has been elusive. Now with JWST, observers have discovered five bound stellar clumps just $460\rm\, Myr$ after the Big Bang at $z\sim10.2^{+0.2}_{-0.2}$ \citep{Adamo24Nat}. These clusters in the strongly lensed galaxy SPT0615-JD1 (alias the Cosmic Gems arc) have intrinsic masses of $\sim 10^6\rm\, M_\odot$, half-light radii of $R_{\rm eff}\sim1\rm\, pc$, and ages between 9 and 35 Myr. Roughly $\sim60\%$ of the total F150W flux of the galaxy comes from these five clusters, indicating that the predominant mode of star formation in these systems occurs in massive clusters. 
The resemblance between M6 and these objects is notable, indicating that the mode of star formation described in this study is a probable path for the formation of YMCs and proto-GCs in the present and early Universe, respectively.
}

\begin{acknowledgements}

B.P. would like to thank Gastón Escobar and Michela Mapelli for the helpful discussions that led to solving a persistent bug allowing continuation of the simulations. \response{B.P. thanks Shyam Menon for many fruitful discussions of YMC formation, particularly regarding the battle between stellar feedback and gravity. We thank the referee for their useful comments and insights.} B.P. was partly supported by a fellowship from the International Max Planck Research School for Astronomy and Cosmic Physics at the University of Heidelberg (IMPRS-HD). M.-M.M.L., B.P., E.P.A., and A.T. were partly supported by NSF grants AST18-15461 and AST23-07950. C.C.-C. is supported by a Canada Graduate Scholarship - Doctoral (CGS D) from the Natural Sciences and Engineering Research Council of Canada (NSERC). This work used Stampede 2 at TACC through allocation PHY220160 from the Advanced Cyberinfrastructure Coordination Ecosystem: Services \& Support (ACCESS) program, which is supported by National Science Foundation grants 21-38259, 21-38286, 21-38307, 21-37603, and 21-38296. The code development that facilitated this study was done on Snellius through the Dutch National Supercomputing Center SURF grants 15220 and 2023/ENW/01498863. S.A. acknowledges the support of NSF grant AST-2009679. M.W. acknowledges the support of NOVA project 10.2.5.12.

R.S.K.\ and S.C.O.G.\ acknowledge financial support from the European Research Council via the ERC Synergy Grant ``ECOGAL'' (project ID 855130),  from the Heidelberg Cluster of Excellence (EXC 2181 - 390900948) ``STRUCTURES'', funded by the German Excellence Strategy, and from the German Ministry for Economic Affairs and Climate Action in project ``MAINN'' (funding ID 50OO2206). The team in Heidelberg also thanks {\em The L\"{a}nd} and the German Science Foundation (DFG) for computing resources provided in bwHPC supported by grant INST 35/1134-1 FUGG and for data storage at SDS@hd supported by grant INST 35/1314-1 FUGG.

L.W. thanks the National Natural Science Foundation of China for support through grants 21BAA00619, 12073090 and 12233013, and the one-hundred-talent project of Sun Yat-sen University, the Fundamental Research Funds for the Central Universities, Sun Yat-sen University (22hytd09).

\response{The authors acknowledge Interstellar Institute's program ``II6'' and the Paris-Saclay University's Institut Pascal for hosting discussions that nourished the development of the ideas behind this work.}

\end{acknowledgements}

\section*{Data Availability}

{\tiny Simulation data and the interactive plot file is available for download at \url{https://doi.org/10.5531/sd.astro.8}.}

\bibliographystyle{aa}
\bibliography{references}

\begin{thebibliography}{159}
\expandafter\ifx\csname natexlab\endcsname\relax\def\natexlab#1{#1}\fi

\bibitem[{Adamo {et~al.}(2024)Adamo, Bradley, Vanzella, Claeyssens, Welch, Diego, Mahler, Oguri, Sharon, Abdurro'uf, Hsiao, Xu, Messa, Lassen, Zackrisson, Brammer, Coe, Kokorev, Ricotti, Zitrin, Fujimoto, Inoue, Resseguier, Rigby, Jim{\'e}nez-Teja, Windhorst, Hashimoto, \& Tamura}]{Adamo24Nat}
Adamo, A., Bradley, L.~D., Vanzella, E., {et~al.} 2024, Nature

\bibitem[{{Adamo} {et~al.}(2023){Adamo}, {Usher}, {Pfeffer}, \& {Claeyssens}}]{Adamo2023MNRAS.525L...6A}
{Adamo}, A., {Usher}, C., {Pfeffer}, J., \& {Claeyssens}, A. 2023, \mnras, 525, L6

\bibitem[{{Adamo} {et~al.}(2020){Adamo}, {Zeidler}, {Kruijssen}, {Chevance}, {Gieles}, {Calzetti}, {Charbonnel}, {Zinnecker}, \& {Krause}}]{Adamo2020SSRv..216...69A}
{Adamo}, A., {Zeidler}, P., {Kruijssen}, J.~M.~D., {et~al.} 2020, \ssr, 216, 69

\bibitem[{{Ali}(2021)}]{Ali2021MNRAS.501.4136A}
{Ali}, A.~A. 2021, \mnras, 501, 4136

\bibitem[{{Andersson} {et~al.}(2024){Andersson}, {Mac Low}, {Agertz}, {Renaud}, \& {Li}}]{Andersson2024A&A...681A..28A}
{Andersson}, E.~P., {Mac Low}, M.-M., {Agertz}, O., {Renaud}, F., \& {Li}, H. 2024, \aap, 681, A28

\bibitem[{{Appel} {et~al.}(2022){Appel}, {Burkhart}, {Semenov}, {Federrath}, \& {Rosen}}]{Appel2022ApJ...927...75A}
{Appel}, S.~M., {Burkhart}, B., {Semenov}, V.~A., {Federrath}, C., \& {Rosen}, A.~L. 2022, \apj, 927, 75

\bibitem[{{Arthur}(2007)}]{Arthur2007ASSP....1..183A}
{Arthur}, S.~J. 2007, in Astrophysics and Space Science Proceedings, Vol.~1, Diffuse Matter from Star Forming Regions to Active Galaxies - A Volume Honouring John Dyson, 183

\bibitem[{{Arthur} {et~al.}(1993){Arthur}, {Dyson}, \& {Hartquist}}]{Arthur1993MNRAS.261..425A}
{Arthur}, S.~J., {Dyson}, J.~E., \& {Hartquist}, T.~W. 1993, \mnras, 261, 425

\bibitem[{{Arthur} {et~al.}(1996){Arthur}, {Henney}, \& {Dyson}}]{Arthur1996A&A...313..897A}
{Arthur}, S.~J., {Henney}, W.~J., \& {Dyson}, J.~E. 1996, \aap, 313, 897

\bibitem[{{Ashman} \& {Zepf}(1992)}]{ashman1992}
{Ashman}, K.~M. \& {Zepf}, S.~E. 1992, \apj, 384, 50

\bibitem[{Baczynski {et~al.}(2015)Baczynski, Glover, \& Klessen}]{FERVENT10.1093/mnras/stv1906}
Baczynski, C., Glover, S. C.~O., \& Klessen, R.~S. 2015, \mnras, 454, 380

\bibitem[{{Ballesteros-Paredes} {et~al.}(2011){Ballesteros-Paredes}, {Hartmann}, {V{\'a}zquez-Semadeni}, {Heitsch}, \& {Zamora-Avil{\'e}s}}]{ballesteros2011}
{Ballesteros-Paredes}, J., {Hartmann}, L.~W., {V{\'a}zquez-Semadeni}, E., {Heitsch}, F., \& {Zamora-Avil{\'e}s}, M.~A. 2011, \mnras, 411, 65

\bibitem[{{Barnes} \& {Hut}(1986)}]{BarnesHut1986Natur.324..446B}
{Barnes}, J. \& {Hut}, P. 1986, \nat, 324, 446

\bibitem[{{Barton} {et~al.}(2000){Barton}, {Geller}, \& {Kenyon}}]{Barton2000ApJ...530..660B}
{Barton}, E.~J., {Geller}, M.~J., \& {Kenyon}, S.~J. 2000, \apj, 530, 660

\bibitem[{{Bate} {et~al.}(1995){Bate}, {Bonnell}, \& {Price}}]{Bate1995MNRAS.277..362B}
{Bate}, M.~R., {Bonnell}, I.~A., \& {Price}, N.~M. 1995, \mnras, 277, 362

\bibitem[{{Bochkarev}(1988)}]{Bochkarev1988Natur.332..518B}
{Bochkarev}, N.~G. 1988, \nat, 332, 518

\bibitem[{{Brodie} \& {Strader}(2006)}]{Brodie2006ARA&A..44..193B}
{Brodie}, J.~P. \& {Strader}, J. 2006, \araa, 44, 193

\bibitem[{{Chevance} {et~al.}(2023){Chevance}, {Krumholz}, {McLeod}, {Ostriker}, {Rosolowsky}, \& {Sternberg}}]{Chevance2023ASPC..534....1C}
{Chevance}, M., {Krumholz}, M.~R., {McLeod}, A.~F., {et~al.} 2023, in Astronomical Society of the Pacific Conference Series, Vol. 534, Protostars and Planets VII, ed. S.~{Inutsuka}, Y.~{Aikawa}, T.~{Muto}, K.~{Tomida}, \& M.~{Tamura}, 1

\bibitem[{{Chu} {et~al.}(2003){Chu}, {Guerrero}, {Gruendl}, {Garc{\'\i}a-Segura}, \& {Wendker}}]{Chu2003ApJ...599.1189C}
{Chu}, Y.-H., {Guerrero}, M.~A., {Gruendl}, R.~A., {Garc{\'\i}a-Segura}, G., \& {Wendker}, H.~J. 2003, \apj, 599, 1189

\bibitem[{{Claeyssens} {et~al.}(2023){Claeyssens}, {Adamo}, {Richard}, {Mahler}, {Messa}, \& {Dessauges-Zavadsky}}]{Claeyssens2023MNRAS.520.2180C}
{Claeyssens}, A., {Adamo}, A., {Richard}, J., {et~al.} 2023, \mnras, 520, 2180

\bibitem[{{Colella} \& {Woodward}(1984)}]{Colella1984JCoPh..54..174C}
{Colella}, P. \& {Woodward}, P.~R. 1984, Journal of Computational Physics, 54, 174

\bibitem[{{Col{\'\i}n} {et~al.}(2013){Col{\'\i}n}, {V{\'a}zquez-Semadeni}, \& {G{\'o}mez}}]{Colin2013MNRAS.435.1701C}
{Col{\'\i}n}, P., {V{\'a}zquez-Semadeni}, E., \& {G{\'o}mez}, G.~C. 2013, \mnras, 435, 1701

\bibitem[{Colín {et~al.}(2013)Colín, Vázquez-Semadeni, \& Gómez}]{10.1093/mnras/stt1409}
Colín, P., Vázquez-Semadeni, E., \& Gómez, G.~C. 2013, \mnras, 435, 1701

\bibitem[{{Cournoyer-Cloutier} {et~al.}(2023){Cournoyer-Cloutier}, {Sills}, {Harris}, {Appel}, {Lewis}, {Polak}, {Tran}, {Wilhelm}, {Mac Low}, {McMillan}, \& {Portegies Zwart}}]{Cournoyer-Cloutier2023MNRAS.521.1338C}
{Cournoyer-Cloutier}, C., {Sills}, A., {Harris}, W.~E., {et~al.} 2023, \mnras, 521, 1338

\bibitem[{{Cournoyer-Cloutier} {et~al.}(2021){Cournoyer-Cloutier}, {Tran}, {Lewis}, {Wall}, {Harris}, {Mac Low}, {McMillan}, {Portegies Zwart}, \& {Sills}}]{2021Cournoyer-Cloutier}
{Cournoyer-Cloutier}, C., {Tran}, A., {Lewis}, S., {et~al.} 2021, MNRAS, 501, 4464

\bibitem[{{Cox} {et~al.}(2008){Cox}, {Jonsson}, {Somerville}, {Primack}, \& {Dekel}}]{Cox2008MNRAS.384..386C}
{Cox}, T.~J., {Jonsson}, P., {Somerville}, R.~S., {Primack}, J.~R., \& {Dekel}, A. 2008, \mnras, 384, 386

\bibitem[{{Crutcher} {et~al.}(2003){Crutcher}, {Heiles}, \& {Troland}}]{Crutcher2003LNP...614..155C}
{Crutcher}, R., {Heiles}, C., \& {Troland}, T. 2003, in Lecture Notes in Physics, Vol. 614, Turbulence and Magnetic Fields in Astrophysics, ed. E.~{Falgarone} \& T.~{Passot} (Berlin, Heidelberg: Springer), 155--181

\bibitem[{{Crutcher} {et~al.}(2010){Crutcher}, {Wandelt}, {Heiles}, {Falgarone}, \& {Troland}}]{Crutcher2010ApJ...725..466C}
{Crutcher}, R.~M., {Wandelt}, B., {Heiles}, C., {Falgarone}, E., \& {Troland}, T.~H. 2010, \apj, 725, 466

\bibitem[{{Dale}(2015)}]{Dale2015NewAR..68....1D}
{Dale}, J.~E. 2015, \nar, 68, 1

\bibitem[{{Dale} {et~al.}(2005){Dale}, {Bonnell}, {Clarke}, \& {Bate}}]{Dale2005MNRAS.358..291D}
{Dale}, J.~E., {Bonnell}, I.~A., {Clarke}, C.~J., \& {Bate}, M.~R. 2005, \mnras, 358, 291

\bibitem[{{Dekel} {et~al.}(2023){Dekel}, {Sarkar}, {Birnboim}, {Mandelker}, \& {Li}}]{Dekel2023MNRAS.523.3201D}
{Dekel}, A., {Sarkar}, K.~C., {Birnboim}, Y., {Mandelker}, N., \& {Li}, Z. 2023, \mnras, 523, 3201

\bibitem[{{Draine}(1978)}]{Draine1978ApJS...36..595D}
{Draine}, B.~T. 1978, \apjs, 36, 595

\bibitem[{Dubey {et~al.}(2014)Dubey, Antypas, Calder, Daley, Fryxell, Gallagher, Lamb, Lee, Olson, Reid, Rich, Ricker, Riley, Rosner, Siegel, Taylor, Weide, Timmes, Vladimirova, \& ZuHone}]{dubey2014}
Dubey, A., Antypas, K., Calder, A.~C., {et~al.} 2014, The International Journal of High Performance Computing Applications, 28, 225

\bibitem[{{Dyson} \& {Hartquist}(1992)}]{Dyson1992ApL&C..28..301D}
{Dyson}, J.~E. \& {Hartquist}, T.~W. 1992, Astrophysical Letters and Communications, 28, 301

\bibitem[{{Ellison} {et~al.}(2008){Ellison}, {Patton}, {Simard}, \& {McConnachie}}]{Ellison2008AJ....135.1877E}
{Ellison}, S.~L., {Patton}, D.~R., {Simard}, L., \& {McConnachie}, A.~W. 2008, \aj, 135, 1877

\bibitem[{{Emig} {et~al.}(2020){Emig}, {Bolatto}, {Leroy}, {Mills}, {Jim{\'e}nez Donaire}, {Tielens}, {Ginsburg}, {Gorski}, {Krieger}, {Levy}, {Meier}, {Ott}, {Rosolowsky}, {Thompson}, \& {Veilleux}}]{Emig2020ApJ...903...50E}
{Emig}, K.~L., {Bolatto}, A.~D., {Leroy}, A.~K., {et~al.} 2020, \apj, 903, 50

\bibitem[{{Evans} {et~al.}(2009){Evans}, {Dunham}, {J{\o}rgensen}, {Enoch}, {Mer{\'\i}n}, {van Dishoeck}, {Alcal{\'a}}, {Myers}, {Stapelfeldt}, {Huard}, {Allen}, {Harvey}, {van Kempen}, {Blake}, {Koerner}, {Mundy}, {Padgett}, \& {Sargent}}]{Evans2009ApJS..181..321E}
{Evans}, Neal~J., I., {Dunham}, M.~M., {J{\o}rgensen}, J.~K., {et~al.} 2009, \apjs, 181, 321

\bibitem[{{Fall}(2006)}]{Fall2006ApJ...652.1129F}
{Fall}, S.~M. 2006, \apj, 652, 1129

\bibitem[{{Fall} {et~al.}(2005){Fall}, {Chandar}, \& {Whitmore}}]{Fall2005ApJ...631L.133F}
{Fall}, S.~M., {Chandar}, R., \& {Whitmore}, B.~C. 2005, \apjl, 631, L133

\bibitem[{{Fall} \& {Zhang}(2001)}]{Fall2001ApJ...561..751F}
{Fall}, S.~M. \& {Zhang}, Q. 2001, \apj, 561, 751

\bibitem[{{Farias} {et~al.}(2023){Farias}, {Offner}, {Grudi{\'c}}, {Guszejnov}, {Rosen}, \& {The Starforge Team}}]{Farias2023MNRAS.tmp.3530F}
{Farias}, J.~P., {Offner}, S. S.~R., {Grudi{\'c}}, M.~Y., {et~al.} 2023, \mnras

\bibitem[{{Federrath}(2015)}]{Fedderath2015MNRAS.450.4035F}
{Federrath}, C. 2015, \mnras, 450, 4035

\bibitem[{{Federrath} {et~al.}(2010){Federrath}, {Banerjee}, {Clark}, \& {Klessen}}]{Federrath2010}
{Federrath}, C., {Banerjee}, R., {Clark}, P.~C., \& {Klessen}, R.~S. 2010, \apj, 713, 269

\bibitem[{Federrath {et~al.}(2010)Federrath, Banerjee, Clark, \& Klessen}]{Federrath_2010}
Federrath, C., Banerjee, R., Clark, P.~C., \& Klessen, R.~S. 2010, \apj, 713, 269

\bibitem[{{Field} {et~al.}(1969){Field}, {Goldsmith}, \& {Habing}}]{field1969}
{Field}, G.~B., {Goldsmith}, D.~W., \& {Habing}, H.~J. 1969, \apjl, 155, L149

\bibitem[{{Finn} {et~al.}(2019){Finn}, {Johnson}, {Brogan}, {Wilson}, {Indebetouw}, {Harris}, {Kamenetzky}, \& {Bemis}}]{Finn2019ApJ...874..120F}
{Finn}, M.~K., {Johnson}, K.~E., {Brogan}, C.~L., {et~al.} 2019, \apj, 874, 120

\bibitem[{{Fryxell} {et~al.}(2000){Fryxell}, {Olson}, {Ricker}, {Timmes}, {Zingale}, {Lamb}, {MacNeice}, {Rosner}, {Truran}, \& {Tufo}}]{flash}
{Fryxell}, B., {Olson}, K., {Ricker}, P., {et~al.} 2000, ApJs, 131, 273

\bibitem[{{Fujii} {et~al.}(2007){Fujii}, {Iwasawa}, {Funato}, \& {Makino}}]{Fujii2007PASJ...59.1095F}
{Fujii}, M., {Iwasawa}, M., {Funato}, Y., \& {Makino}, J. 2007, \pasj, 59, 1095

\bibitem[{{Fujii} {et~al.}(2022{\natexlab{a}}){Fujii}, {Hattori}, {Wang}, {Hirai}, {Kumamoto}, {Shimajiri}, \& {Saitoh}}]{Fujii2022b}
{Fujii}, M.~S., {Hattori}, K., {Wang}, L., {et~al.} 2022{\natexlab{a}}, \mnras, 514, 43

\bibitem[{{Fujii} \& {Portegies Zwart}(2011)}]{Fujii2011Sci...334.1380F}
{Fujii}, M.~S. \& {Portegies Zwart}, S. 2011, Science, 334, 1380

\bibitem[{{Fujii} {et~al.}(2021){Fujii}, {Saitoh}, {Hirai}, \& {Wang}}]{Fujii2021PASJ...73.1074F}
{Fujii}, M.~S., {Saitoh}, T.~R., {Hirai}, Y., \& {Wang}, L. 2021, \pasj, 73, 1074

\bibitem[{{Fujii} {et~al.}(2022{\natexlab{b}}){Fujii}, {Wang}, {Hirai}, {Shimajiri}, {Kumamoto}, \& {Saitoh}}]{Fujii2022a}
{Fujii}, M.~S., {Wang}, L., {Hirai}, Y., {et~al.} 2022{\natexlab{b}}, \mnras, 514, 2513

\bibitem[{{Fujii} {et~al.}(2024){Fujii}, {Wang}, {Tanikawa}, {Hirai}, \& {Saitoh}}]{Fujii2024arXiv240606772F}
{Fujii}, M.~S., {Wang}, L., {Tanikawa}, A., {Hirai}, Y., \& {Saitoh}, T.~R. 2024, arXiv e-prints, arXiv:2406.06772

\bibitem[{{Fukushima} \& {Yajima}(2021)}]{Fukushima2021MNRAS.506.5512F}
{Fukushima}, H. \& {Yajima}, H. 2021, \mnras, 506, 5512

\bibitem[{{Galv{\'a}n-Madrid} {et~al.}(2013){Galv{\'a}n-Madrid}, {Liu}, {Zhang}, {Pineda}, {Peng}, {Zhang}, {Keto}, {Ho}, {Rodr{\'\i}guez}, {Zapata}, {Peters}, \& {De Pree}}]{GalvanMadrid2013ApJ...779..121G}
{Galv{\'a}n-Madrid}, R., {Liu}, H.~B., {Zhang}, Z.~Y., {et~al.} 2013, \apj, 779, 121

\bibitem[{{Geen} {et~al.}(2017){Geen}, {Soler}, \& {Hennebelle}}]{Geen2017MNRAS.471.4844G}
{Geen}, S., {Soler}, J.~D., \& {Hennebelle}, P. 2017, \mnras, 471, 4844

\bibitem[{{Geyer} \& {Burkert}(2001)}]{Geyer2001MNRAS.323..988G}
{Geyer}, M.~P. \& {Burkert}, A. 2001, \mnras, 323, 988

\bibitem[{{Girichidis} {et~al.}(2020){Girichidis}, {Offner}, {Kritsuk}, {Klessen}, {Hennebelle}, {Kruijssen}, {Krause}, {Glover}, \& {Padovani}}]{Girichidis2020SSRv..216...68G}
{Girichidis}, P., {Offner}, S. S.~R., {Kritsuk}, A.~G., {et~al.} 2020, \ssr, 216, 68

\bibitem[{{Goodwin} {et~al.}(2004){Goodwin}, {Whitworth}, \& {Ward-Thompson}}]{Goodwin2004A&A...414..633G}
{Goodwin}, S.~P., {Whitworth}, A.~P., \& {Ward-Thompson}, D. 2004, \aap, 414, 633

\bibitem[{{Grudi{\'c}} {et~al.}(2021){Grudi{\'c}}, {Guszejnov}, {Hopkins}, {Offner}, \& {Faucher-Gigu{\`e}re}}]{2021MNRAS.506.2199G}
{Grudi{\'c}}, M.~Y., {Guszejnov}, D., {Hopkins}, P.~F., {Offner}, S. S.~R., \& {Faucher-Gigu{\`e}re}, C.-A. 2021, \mnras, 506, 2199

\bibitem[{{Grudi{\'c}} {et~al.}(2018){Grudi{\'c}}, {Hopkins}, {Faucher-Gigu{\`e}re}, {Quataert}, {Murray}, \& {Kere{\v{s}}}}]{Grudic2018MNRAS.475.3511G}
{Grudi{\'c}}, M.~Y., {Hopkins}, P.~F., {Faucher-Gigu{\`e}re}, C.-A., {et~al.} 2018, \mnras, 475, 3511

\bibitem[{{Grudi{\'c}} {et~al.}(2019){Grudi{\'c}}, {Hopkins}, {Lee}, {Murray}, {Faucher-Gigu{\`e}re}, \& {Johnson}}]{Grudic2019MNRAS.488.1501G}
{Grudi{\'c}}, M.~Y., {Hopkins}, P.~F., {Lee}, E.~J., {et~al.} 2019, \mnras, 488, 1501

\bibitem[{Grudić {et~al.}(2018)Grudić, Hopkins, Faucher-Giguère, Quataert, Murray, \& Kereš}]{10.1093/mnras/sty035}
Grudić, M.~Y., Hopkins, P.~F., Faucher-Giguère, C.-A., {et~al.} 2018, \mnras, 475, 3511

\bibitem[{{Guszejnov} {et~al.}(2021){Guszejnov}, {Grudi{\'c}}, {Hopkins}, {Offner}, \& {Faucher-Gigu{\`e}re}}]{Guszejnov2021MNRAS.502.3646G}
{Guszejnov}, D., {Grudi{\'c}}, M.~Y., {Hopkins}, P.~F., {Offner}, S. S.~R., \& {Faucher-Gigu{\`e}re}, C.-A. 2021, \mnras, 502, 3646

\bibitem[{{Guszejnov} {et~al.}(2022){Guszejnov}, {Grudi{\'c}}, {Offner}, {Faucher-Gigu{\`e}re}, {Hopkins}, \& {Rosen}}]{guszejnov2022}
{Guszejnov}, D., {Grudi{\'c}}, M.~Y., {Offner}, S. S.~R., {et~al.} 2022, \mnras, 515, 4929

\bibitem[{{Harris} \& {Reina-Campos}(2023)}]{Harris2023MNRAS.526.2696H}
{Harris}, W.~E. \& {Reina-Campos}, M. 2023, \mnras, 526, 2696

\bibitem[{{Hartquist} \& {Dyson}(1996)}]{Hartquist1996Ap&SS.245..263H}
{Hartquist}, T.~W. \& {Dyson}, J.~E. 1996, \apss, 245, 263

\bibitem[{{Hartquist} {et~al.}(1986){Hartquist}, {Dyson}, {Pettini}, \& {Smith}}]{Hartquist1986MNRAS.221..715H}
{Hartquist}, T.~W., {Dyson}, J.~E., {Pettini}, M., \& {Smith}, L.~J. 1986, \mnras, 221, 715

\bibitem[{{He} {et~al.}(2019){He}, {Ricotti}, \& {Geen}}]{He2019MNRAS.489.1880H}
{He}, C.-C., {Ricotti}, M., \& {Geen}, S. 2019, \mnras, 489, 1880

\bibitem[{Heiles(1976)}]{Heiles76}
Heiles, C. 1976, \araa, 14, 1

\bibitem[{{Heitsch} {et~al.}(2001){Heitsch}, {Mac Low}, \& {Klessen}}]{Heitsch2001ApJ...547..280H}
{Heitsch}, F., {Mac Low}, M.-M., \& {Klessen}, R.~S. 2001, \apj, 547, 280

\bibitem[{Howard {et~al.}(2017)Howard, Pudritz, \& Harris}]{10.1093/mnras/stx1363}
Howard, C.~S., Pudritz, R.~E., \& Harris, W.~E. 2017, \mnras, 470, 3346

\bibitem[{{Howard} {et~al.}(2018){Howard}, {Pudritz}, \& {Harris}}]{Howard2018NatAs...2..725H}
{Howard}, C.~S., {Pudritz}, R.~E., \& {Harris}, W.~E. 2018, Nature Astronomy, 2, 725

\bibitem[{{Iwasawa} {et~al.}(2020){Iwasawa}, {Namekata}, {Nitadori}, {Nomura}, {Wang}, {Tsubouchi}, \& {Makino}}]{Iwasawa2020}
{Iwasawa}, M., {Namekata}, D., {Nitadori}, K., {et~al.} 2020, \pasj, 72, 13

\bibitem[{{Iwasawa} {et~al.}(2016){Iwasawa}, {Tanikawa}, {Hosono}, {Nitadori}, {Muranushi}, \& {Makino}}]{Iwasawa2016}
{Iwasawa}, M., {Tanikawa}, A., {Hosono}, N., {et~al.} 2016, \pasj, 68, 54

\bibitem[{{Johnson} {et~al.}(2015){Johnson}, {Leroy}, {Indebetouw}, {Brogan}, {Whitmore}, {Hibbard}, {Sheth}, \& {Evans}}]{Johnson2015ApJ...806...35J}
{Johnson}, K.~E., {Leroy}, A.~K., {Indebetouw}, R., {et~al.} 2015, \apj, 806, 35

\bibitem[{{Kauffmann} {et~al.}(2013){Kauffmann}, {Pillai}, \& {Goldsmith}}]{2013ApJ...779..185K}
{Kauffmann}, J., {Pillai}, T., \& {Goldsmith}, P.~F. 2013, ApJ, 779, 185

\bibitem[{{Kim} {et~al.}(2018){Kim}, {Kim}, \& {Ostriker}}]{Kim2018ApJ...859...68K}
{Kim}, J.-G., {Kim}, W.-T., \& {Ostriker}, E.~C. 2018, \apj, 859, 68

\bibitem[{Kim {et~al.}(2017)Kim, Kim, Ostriker, \& Skinner}]{Kim_2017}
Kim, J.-G., Kim, W.-T., Ostriker, E.~C., \& Skinner, M.~A. 2017, \apj, 851, 93

\bibitem[{{Kimm} {et~al.}(2022){Kimm}, {Bieri}, {Geen}, {Rosdahl}, {Blaizot}, {Michel-Dansac}, \& {Garel}}]{Kimm2022ApJS..259...21K}
{Kimm}, T., {Bieri}, R., {Geen}, S., {et~al.} 2022, \apjs, 259, 21

\bibitem[{{Klessen} \& {Glover}(2016)}]{Klessen2016SAAS...43...85K}
{Klessen}, R.~S. \& {Glover}, S. C.~O. 2016, in Saas-Fee Advanced Course, Vol.~43, Saas-Fee Advanced Course, ed. Y.~{Revaz}, P.~{Jablonka}, R.~{Teyssier}, \& L.~{Mayer}, 85

\bibitem[{{Kolmogorov}(1941)}]{Kolmogorov1941DoSSR..30..301K}
{Kolmogorov}, A. 1941, Akademiia Nauk SSSR Doklady, 30, 301

\bibitem[{{Krause} {et~al.}(2020){Krause}, {Offner}, {Charbonnel}, {Gieles}, {Klessen}, {V{\'a}zquez-Semadeni}, {Ballesteros-Paredes}, {Girichidis}, {Kruijssen}, {Ward}, \& {Zinnecker}}]{Krause2020SSRv..216...64K}
{Krause}, M. G.~H., {Offner}, S. S.~R., {Charbonnel}, C., {et~al.} 2020, \ssr, 216, 64

\bibitem[{{Kroupa}(2002)}]{2002Sci...295...82Kroupa}
{Kroupa}, P. 2002, Science, 295, 82

\bibitem[{{Krumholz} {et~al.}(2019){Krumholz}, {McKee}, \& {Bland-Hawthorn}}]{Krumholz2019ARA&A..57..227K}
{Krumholz}, M.~R., {McKee}, C.~F., \& {Bland-Hawthorn}, J. 2019, \araa, 57, 227

\bibitem[{{Kudritzki} \& {Puls}(2000)}]{Kudritzki2000ARA&A..38..613K}
{Kudritzki}, R.-P. \& {Puls}, J. 2000, \araa, 38, 613

\bibitem[{{Lada} \& {Lada}(2003)}]{Lada2003ARA&A..41...57L}
{Lada}, C.~J. \& {Lada}, E.~A. 2003, \araa, 41, 57

\bibitem[{{Lah{\'e}n} {et~al.}(2019){Lah{\'e}n}, {Naab}, {Johansson}, {Elmegreen}, {Hu}, \& {Walch}}]{Lahen2019ApJ...879L..18L}
{Lah{\'e}n}, N., {Naab}, T., {Johansson}, P.~H., {et~al.} 2019, \apjl, 879, L18

\bibitem[{{Lah{\'e}n} {et~al.}(2024){Lah{\'e}n}, {Naab}, \& {Sz{\'e}csi}}]{Lahen2024MNRAS.530..645L}
{Lah{\'e}n}, N., {Naab}, T., \& {Sz{\'e}csi}, D. 2024, \mnras, 530, 645

\bibitem[{{Lancaster} {et~al.}(2021){Lancaster}, {Ostriker}, {Kim}, \& {Kim}}]{Lancaster2021ApJ...914...89L}
{Lancaster}, L., {Ostriker}, E.~C., {Kim}, J.-G., \& {Kim}, C.-G. 2021, \apj, 914, 89

\bibitem[{{Larson}(1981)}]{Larson1981MNRAS.194..809L}
{Larson}, R.~B. 1981, \mnras, 194, 809

\bibitem[{{Larson} \& {Tinsley}(1978)}]{Larson1978ApJ...219...46L}
{Larson}, R.~B. \& {Tinsley}, B.~M. 1978, \apj, 219, 46

\bibitem[{{Lee} {et~al.}(2016){Lee}, {Miville-Desch{\^e}nes}, \& {Murray}}]{Lee2016ApJ...833..229L}
{Lee}, E.~J., {Miville-Desch{\^e}nes}, M.-A., \& {Murray}, N.~W. 2016, \apj, 833, 229

\bibitem[{{Leroy} {et~al.}(2018){Leroy}, {Bolatto}, {Ostriker}, {Walter}, {Gorski}, {Ginsburg}, {Krieger}, {Levy}, {Meier}, {Mills}, {Ott}, {Rosolowsky}, {Thompson}, {Veilleux}, \& {Zschaechner}}]{Leroy2018ApJ...869..126L}
{Leroy}, A.~K., {Bolatto}, A.~D., {Ostriker}, E.~C., {et~al.} 2018, \apj, 869, 126

\bibitem[{{Lewis} {et~al.}(2023){Lewis}, {McMillan}, {Low}, {Cournoyer-Cloutier}, {Polak}, {Wilhelm}, {Tran}, {Sills}, {Portegies Zwart}, {Klessen}, \& {Wall}}]{Lewis2023ApJ...944..211L}
{Lewis}, S.~C., {McMillan}, S. L.~W., {Low}, M.-M.~M., {et~al.} 2023, \apj, 944, 211

\bibitem[{{Li} {et~al.}(2018){Li}, {Gnedin}, \& {Gnedin}}]{Li2018ApJ...861..107L}
{Li}, H., {Gnedin}, O.~Y., \& {Gnedin}, N.~Y. 2018, \apj, 861, 107

\bibitem[{Li {et~al.}(2019)Li, Vogelsberger, Marinacci, \& Gnedin}]{10.1093/mnras/stz1271}
Li, H., Vogelsberger, M., Marinacci, F., \& Gnedin, O.~Y. 2019, \mnras, 487, 364

\bibitem[{{Li} {et~al.}(2024){Li}, {Dekel}, {Sarkar}, {Aung}, {Giavalisco}, {Mandelker}, \& {Tacchella}}]{Li2023arXiv231114662L}
{Li}, Z., {Dekel}, A., {Sarkar}, K.~C., {et~al.} 2024, \aap, accepted

\bibitem[{{Lin} {et~al.}(2016){Lin}, {Liu}, {Li}, {Zhang}, {Ginsburg}, {Pineda}, {Qian}, {Galv{\'a}n-Madrid}, {McLeod}, {Rosolowsky}, {Dale}, {Immer}, {Koch}, {Longmore}, {Walker}, \& {Testi}}]{Lin2016ApJ...828...32L}
{Lin}, Y., {Liu}, H.~B., {Li}, D., {et~al.} 2016, \apj, 828, 32

\bibitem[{{L{\"o}hner}(1987)}]{Lohner1987CMAME..61..323L}
{L{\"o}hner}, R. 1987, Computer Methods in Applied Mechanics and Engineering, 61, 323

\bibitem[{{Lonsdale} {et~al.}(1984){Lonsdale}, {Persson}, \& {Matthews}}]{Lonsdale1984ApJ...287...95L}
{Lonsdale}, C.~J., {Persson}, S.~E., \& {Matthews}, K. 1984, \apj, 287, 95

\bibitem[{Mac~Low \& Klessen(2004)}]{MacLowRevModPhys.76.125}
Mac~Low, M.-M. \& Klessen, R.~S. 2004, Rev. Mod. Phys., 76, 125

\bibitem[{{Makino} \& {Aarseth}(1992)}]{Makino1992PASJ...44..141M}
{Makino}, J. \& {Aarseth}, S.~J. 1992, \pasj, 44, 141

\bibitem[{{Matzner}(2002)}]{matzner2002}
{Matzner}, C.~D. 2002, \apj, 566, 302

\bibitem[{McKee \& Ostriker(2007)}]{Mckeedoi:10.1146/annurev.astro.45.051806.110602}
McKee, C.~F. \& Ostriker, E.~C. 2007, \araa, 45, 565

\bibitem[{McKee \& Williams(1997)}]{McKee_1997}
McKee, C.~F. \& Williams, J.~P. 1997, \apj, 476, 144

\bibitem[{{McMillan} {et~al.}(2012){McMillan}, {Portegies Zwart}, {van Elteren}, \& {Whitehead}}]{ph4}
{McMillan}, S., {Portegies Zwart}, S., {van Elteren}, A., \& {Whitehead}, A. 2012, in Astronomical Society of the Pacific Conference Series, Vol. 453, Advances in Computational Astrophysics: Methods, Tools, and Outcome, ed. R.~{Capuzzo-Dolcetta}, M.~{Limongi}, \& A.~{Tornamb{\`e}}, 129

\bibitem[{{Menon} {et~al.}(2023){Menon}, {Federrath}, \& {Krumholz}}]{Menon2023MNRAS.521.5160M}
{Menon}, S.~H., {Federrath}, C., \& {Krumholz}, M.~R. 2023, \mnras, 521, 5160

\bibitem[{{Miyoshi} \& {Kusano}(2005)}]{Miyoshi2005JCoPh.208..315M}
{Miyoshi}, T. \& {Kusano}, K. 2005, Journal of Computational Physics, 208, 315

\bibitem[{Mouschovias(1991)}]{Mouschovias1991}
Mouschovias, T.~C. 1991, Cosmic Magnetism and the Basic Physics of the Early Stages of Star Formation (Dordrecht: Springer Netherlands), 61--122

\bibitem[{{Mouschovias} \& {Spitzer}(1976)}]{1976Mouschovias}
{Mouschovias}, T.~C. \& {Spitzer}, L., J. 1976, ApJ, 210, 326

\bibitem[{Murray \& Rahman(2009)}]{Murray_2010}
Murray, N. \& Rahman, M. 2009, \apj, 709, 424

\bibitem[{{Nakamura} \& {Li}(2007)}]{Nakamura2007ApJ...662..395N}
{Nakamura}, F. \& {Li}, Z.-Y. 2007, \apj, 662, 395

\bibitem[{{Peters} {et~al.}(2011){Peters}, {Banerjee}, {Klessen}, \& {Mac Low}}]{Peters2011ApJ...729...72P}
{Peters}, T., {Banerjee}, R., {Klessen}, R.~S., \& {Mac Low}, M.-M. 2011, \apj, 729, 72

\bibitem[{{Pittard} {et~al.}(2001){Pittard}, {Hartquist}, \& {Dyson}}]{Pittard2001A&A...373.1043P}
{Pittard}, J.~M., {Hartquist}, T.~W., \& {Dyson}, J.~E. 2001, \aap, 373, 1043

\bibitem[{{Plotly Technologies Inc.}(2015)}]{plotly}
{Plotly Technologies Inc.} 2015, Collaborative data science

\bibitem[{Portegies~Zwart \& McMillan(2018)}]{amusebook}
Portegies~Zwart, S. \& McMillan, S. 2018, Astrophysical Recipes, 2514-3433 (IOP Publishing)

\bibitem[{{Portegies Zwart} {et~al.}(2009){Portegies Zwart}, McMillan, Harfst, Groen, Fujii, NuallÃ¡in, Glebbeek, Heggie, Lombardi, Hut, Angelou, Banerjee, Belkus, Fragos, Fregeau, Gaburov, Izzard, JuriÄ, Justham, Sottoriva, Teuben, {van Bever}, Yaron, \& Zemp}]{PORTEGIESZWART2009369amuse1}
{Portegies Zwart}, S., McMillan, S., Harfst, S., {et~al.} 2009, New Astronomy, 14, 369

\bibitem[{{Portegies Zwart} {et~al.}(2010){Portegies Zwart}, {McMillan}, \& {Gieles}}]{PortegiesZwart2010ARA&A..48..431P}
{Portegies Zwart}, S.~F., {McMillan}, S. L.~W., \& {Gieles}, M. 2010, \araa, 48, 431

\bibitem[{{Portegies Zwart} \& {Verbunt}(1996)}]{SeBa}
{Portegies Zwart}, S.~F. \& {Verbunt}, F. 1996, \aap, 309, 179

\bibitem[{{Price} \& {Bate}(2008)}]{Price2008MNRAS.385.1820P}
{Price}, D.~J. \& {Bate}, M.~R. 2008, \mnras, 385, 1820

\bibitem[{{Rahner} {et~al.}(2019){Rahner}, {Pellegrini}, {Glover}, \& {Klessen}}]{Rahner2019MNRAS.483.2547R}
{Rahner}, D., {Pellegrini}, E.~W., {Glover}, S. C.~O., \& {Klessen}, R.~S. 2019, \mnras, 483, 2547

\bibitem[{{Reina-Campos} \& {Harris}(2024)}]{ReinaCampos2024MNRAS.531.4099R}
{Reina-Campos}, M. \& {Harris}, W.~E. 2024, \mnras, 531, 4099

\bibitem[{{Renaud}(2020)}]{Renaud2020IAUS..351...40R}
{Renaud}, F. 2020, in Star Clusters: From the Milky Way to the Early Universe, ed. A.~{Bragaglia}, M.~{Davies}, A.~{Sills}, \& E.~{Vesperini}, Vol. 351, 40--46

\bibitem[{{Renaud} {et~al.}(2017){Renaud}, {Agertz}, \& {Gieles}}]{Renaud2017MNRAS.465.3622R}
{Renaud}, F., {Agertz}, O., \& {Gieles}, M. 2017, \mnras, 465, 3622

\bibitem[{{Renaud} {et~al.}(2019){Renaud}, {Bournaud}, {Agertz}, {Kraljic}, {Schinnerer}, {Bolatto}, {Daddi}, \& {Hughes}}]{Renaud2019A&A...625A..65R}
{Renaud}, F., {Bournaud}, F., {Agertz}, O., {et~al.} 2019, \aap, 625, A65

\bibitem[{{Rice} {et~al.}(2016){Rice}, {Goodman}, {Bergin}, {Beaumont}, \& {Dame}}]{Rice2016ApJ...822...52R}
{Rice}, T.~S., {Goodman}, A.~A., {Bergin}, E.~A., {Beaumont}, C., \& {Dame}, T.~M. 2016, \apj, 822, 52

\bibitem[{{Rico-Villas} {et~al.}(2020){Rico-Villas}, {Mart{\'\i}n-Pintado}, {Gonz{\'a}lez-Alfonso}, {Mart{\'\i}n}, \& {Rivilla}}]{Rico-Villas2020MNRAS.491.4573R}
{Rico-Villas}, F., {Mart{\'\i}n-Pintado}, J., {Gonz{\'a}lez-Alfonso}, E., {Mart{\'\i}n}, S., \& {Rivilla}, V.~M. 2020, \mnras, 491, 4573

\bibitem[{{Simpson} {et~al.}(2015){Simpson}, {Bryan}, {Hummels}, \& {Ostriker}}]{Simpson2015ApJ...809...69S}
{Simpson}, C.~M., {Bryan}, G.~L., {Hummels}, C., \& {Ostriker}, J.~P. 2015, \apj, 809, 69

\bibitem[{{Smith} {et~al.}(1984){Smith}, {Pettini}, {Dyson}, \& {Hartquist}}]{Smith1984MNRAS.211..679S}
{Smith}, L.~J., {Pettini}, M., {Dyson}, J.~E., \& {Hartquist}, T.~W. 1984, \mnras, 211, 679

\bibitem[{{Sormani} {et~al.}(2017){Sormani}, {Tre{\ss}}, {Klessen}, \& {Glover}}]{Sormani2017}
{Sormani}, M.~C., {Tre{\ss}}, R.~G., {Klessen}, R.~S., \& {Glover}, S. C.~O. 2017, \mnras, 466, 407

\bibitem[{{Strittmatter}(1966)}]{1966Strittmatter}
{Strittmatter}, P.~A. 1966, MNRAS, 132, 359

\bibitem[{Su {et~al.}(2018)Su, Hopkins, Hayward, Ma, Boylan-Kolchin, Kasen, Kereš, Faucher-Giguère, Orr, \& Wheeler}]{10.1093/mnras/sty1928}
Su, K.-Y., Hopkins, P.~F., Hayward, C.~C., {et~al.} 2018, \mnras, 480, 1666

\bibitem[{{Sun} {et~al.}(2022){Sun}, {Leroy}, {Rosolowsky}, {Hughes}, {Schinnerer}, {Schruba}, {Koch}, {Blanc}, {Chiang}, {Groves}, {Liu}, {Meidt}, {Pan}, {Pety}, {Querejeta}, {Saito}, {Sandstrom}, {Sardone}, {Usero}, {Utomo}, {Williams}, {Barnes}, {Benincasa}, {Bigiel}, {Bolatto}, {Boquien}, {Chevance}, {Dale}, {Deger}, {Emsellem}, {Glover}, {Grasha}, {Henshaw}, {Klessen}, {Kreckel}, {Kruijssen}, {Ostriker}, \& {Thilker}}]{Sun2022AJ....164...43S}
{Sun}, J., {Leroy}, A.~K., {Rosolowsky}, E., {et~al.} 2022, \aj, 164, 43

\bibitem[{{Tacconi} {et~al.}(2020){Tacconi}, {Genzel}, \& {Sternberg}}]{Tacconi2020ARA&A..58..157T}
{Tacconi}, L.~J., {Genzel}, R., \& {Sternberg}, A. 2020, \araa, 58, 157

\bibitem[{{Tress} {et~al.}(2020){Tress}, {Smith}, {Sormani}, {Glover}, {Klessen}, {Mac Low}, \& {Clark}}]{Tress2020MNRAS.492.2973T}
{Tress}, R.~G., {Smith}, R.~J., {Sormani}, M.~C., {et~al.} 2020, \mnras, 492, 2973

\bibitem[{{Truelove} {et~al.}(1997){Truelove}, {Klein}, {McKee}, {Holliman}, {Howell}, \& {Greenough}}]{Truelove1997ApJ...489L.179T}
{Truelove}, J.~K., {Klein}, R.~I., {McKee}, C.~F., {et~al.} 1997, \apjl, 489, L179

\bibitem[{{Turner} {et~al.}(2015){Turner}, {Beck}, {Benford}, {Consiglio}, {Ho}, {Kov{\'a}cs}, {Meier}, \& {Zhao}}]{Turner2015Natur.519..331T}
{Turner}, J.~L., {Beck}, S.~C., {Benford}, D.~J., {et~al.} 2015, \nat, 519, 331

\bibitem[{{van den Bergh}(2001)}]{vandenBergh2001ApJ...559L.113V}
{van den Bergh}, S. 2001, \apjl, 559, L113

\bibitem[{{Vanzella} {et~al.}(2019){Vanzella}, {Calura}, {Meneghetti}, {Castellano}, {Caminha}, {Mercurio}, {Cupani}, {Rosati}, {Grillo}, {Gilli}, {Mignoli}, {Fiorentino}, {Arcidiacono}, {Lombini}, \& {Cortecchia}}]{Vanzella2019MNRAS.483.3618V}
{Vanzella}, E., {Calura}, F., {Meneghetti}, M., {et~al.} 2019, \mnras, 483, 3618

\bibitem[{{Vanzella} {et~al.}(2022{\natexlab{a}}){Vanzella}, {Castellano}, {Bergamini}, {Meneghetti}, {Zanella}, {Calura}, {Caminha}, {Rosati}, {Cupani}, {Me{\v{s}}tri{\'c}}, {Brammer}, {Tozzi}, {Mercurio}, {Grillo}, {Sani}, {Cristiani}, {Nonino}, {Merlin}, \& {Pignataro}}]{Vanzella2022A&A...659A...2V}
{Vanzella}, E., {Castellano}, M., {Bergamini}, P., {et~al.} 2022{\natexlab{a}}, \aap, 659, A2

\bibitem[{{Vanzella} {et~al.}(2022{\natexlab{b}}){Vanzella}, {Castellano}, {Bergamini}, {Treu}, {Mercurio}, {Scarlata}, {Rosati}, {Grillo}, {Acebron}, {Caminha}, {Nonino}, {Nanayakkara}, {Roberts-Borsani}, {Bradac}, {Wang}, {Brammer}, {Strait}, {Vulcani}, {Me{\v{s}}tri{\'c}}, {Meneghetti}, {Calura}, {Henry}, {Zanella}, {Trenti}, {Boyett}, {Morishita}, {Calabr{\`o}}, {Glazebrook}, {Marchesini}, {Birrer}, {Yang}, \& {Jones}}]{Vanzella2022ApJ...940L..53V}
{Vanzella}, E., {Castellano}, M., {Bergamini}, P., {et~al.} 2022{\natexlab{b}}, \apjl, 940, L53

\bibitem[{{Vanzella} {et~al.}(2023){Vanzella}, {Claeyssens}, {Welch}, {Adamo}, {Coe}, {Diego}, {Mahler}, {Khullar}, {Kokorev}, {Oguri}, {Ravindranath}, {Furtak}, {Hsiao}, {Abdurro'uf}, {Mandelker}, {Brammer}, {Bradley}, {Brada{\v{c}}}, {Conselice}, {Dayal}, {Nonino}, {Andrade-Santos}, {Windhorst}, {Pirzkal}, {Sharon}, {de Mink}, {Fujimoto}, {Zitrin}, {Eldridge}, \& {Norman}}]{Vanzella2023ApJ...945...53V}
{Vanzella}, E., {Claeyssens}, A., {Welch}, B., {et~al.} 2023, \apj, 945, 53

\bibitem[{{Vink} {et~al.}(2000){Vink}, {de Koter}, \& {Lamers}}]{Vink2000A&A...362..295V}
{Vink}, J.~S., {de Koter}, A., \& {Lamers}, H.~J.~G.~L.~M. 2000, \aap, 362, 295

\bibitem[{{Wall} {et~al.}(2020){Wall}, {Mac Low}, {McMillan}, {Klessen}, {Portegies Zwart}, \& {Pellegrino}}]{2020Wall}
{Wall}, J.~E., {Mac Low}, M.-M., {McMillan}, S. L.~W., {et~al.} 2020, ApJ, 904, 192

\bibitem[{{Wall} {et~al.}(2019){Wall}, {McMillan}, {Mac Low}, {Klessen}, \& {Portegies Zwart}}]{wall2019}
{Wall}, J.~E., {McMillan}, S. L.~W., {Mac Low}, M.-M., {Klessen}, R.~S., \& {Portegies Zwart}, S. 2019, ApJ, 887, 62

\bibitem[{{Wang} {et~al.}(2020{\natexlab{a}}){Wang}, {Iwasawa}, {Nitadori}, \& {Makino}}]{PeTar}
{Wang}, L., {Iwasawa}, M., {Nitadori}, K., \& {Makino}, J. 2020{\natexlab{a}}, MNRAS, 497, 536

\bibitem[{{Wang} {et~al.}(2020{\natexlab{b}}){Wang}, {Nitadori}, \& {Makino}}]{SDAR2020MNRAS.493.3398W}
{Wang}, L., {Nitadori}, K., \& {Makino}, J. 2020{\natexlab{b}}, \mnras, 493, 3398

\bibitem[{{Weaver} {et~al.}(1977){Weaver}, {McCray}, {Castor}, {Shapiro}, \& {Moore}}]{Weaver1977ApJ...218..377W}
{Weaver}, R., {McCray}, R., {Castor}, J., {Shapiro}, P., \& {Moore}, R. 1977, \apj, 218, 377

\bibitem[{{Wei} {et~al.}(2012){Wei}, {Keto}, \& {Ho}}]{Wei2012ApJ...750..136W}
{Wei}, L.~H., {Keto}, E., \& {Ho}, L.~C. 2012, \apj, 750, 136

\bibitem[{Weidner {et~al.}(2009)Weidner, Kroupa, \& Bonnell}]{Weidner10.1111/j.1365-2966.2009.15633.x}
Weidner, C., Kroupa, P., \& Bonnell, I. A.~D. 2009, \mnras, 401, 275

\bibitem[{{Whitmore} {et~al.}(2014){Whitmore}, {Brogan}, {Chandar}, {Evans}, {Hibbard}, {Johnson}, {Leroy}, {Privon}, {Remijan}, \& {Sheth}}]{Whitmore2014ApJ...795..156W}
{Whitmore}, B.~C., {Brogan}, C., {Chandar}, R., {et~al.} 2014, \apj, 795, 156

\bibitem[{{Whitmore} {et~al.}(2023){Whitmore}, {Chandar}, {Rodr{\'\i}guez}, {Lee}, {Emsellem}, {Floyd}, {Kim}, {Kruijssen}, {Mok}, {Sormani}, {Boquien}, {Dale}, {Faesi}, {Henny}, {Hannon}, {Thilker}, {White}, {Barnes}, {Bigiel}, {Chevance}, {Henshaw}, {Klessen}, {Leroy}, {Liu}, {Maschmann}, {Meidt}, {Rosolowsky}, {Schinnerer}, {Sun}, {Watkins}, \& {Williams}}]{Whitmore2023ApJ...944L..14W}
{Whitmore}, B.~C., {Chandar}, R., {Rodr{\'\i}guez}, M.~J., {et~al.} 2023, \apjl, 944, L14

\bibitem[{{Wilhelm} {et~al.}(2023){Wilhelm}, {Portegies Zwart}, {Cournoyer-Cloutier}, {Lewis}, {Polak}, {Tran}, \& {Mac Low}}]{Wilhelm2023MNRAS.520.5331W}
{Wilhelm}, M. J.~C., {Portegies Zwart}, S., {Cournoyer-Cloutier}, C., {et~al.} 2023, \mnras, 520, 5331

\bibitem[{{Wolfire} {et~al.}(2003){Wolfire}, {McKee}, {Hollenbach}, \& {Tielens}}]{wolfire2003}
{Wolfire}, M.~G., {McKee}, C.~F., {Hollenbach}, D., \& {Tielens}, A.~G.~G.~M. 2003, \apj, 587, 278

\bibitem[{{Wrigge}(1999)}]{Wrigge1999A&A...343..599W}
{Wrigge}, M. 1999, \aap, 343, 599

\bibitem[{{Wrigge} {et~al.}(1994){Wrigge}, {Wendker}, \& {Wisotzki}}]{Wrigge1994A&A...286..219W}
{Wrigge}, M., {Wendker}, H.~J., \& {Wisotzki}, L. 1994, \aap, 286, 219

\bibitem[{{Yan} {et~al.}(2023){Yan}, {Jerabkova}, \& {Kroupa}}]{Yan2023A&A...670A.151Y}
{Yan}, Z., {Jerabkova}, T., \& {Kroupa}, P. 2023, \aap, 670, A151

\bibitem[{{Zamora-Avil{\'e}s} \& {V{\'a}zquez-Semadeni}(2014)}]{Zamora-Aviles2014ApJ...793...84Z}
{Zamora-Avil{\'e}s}, M. \& {V{\'a}zquez-Semadeni}, E. 2014, \apj, 793, 84

\end{thebibliography}

\begin{appendix}

\section{Stellar properties: Physical times}
\label{app:props_physical}

\begin{figure*}
\centering
    \includegraphics[width=1\textwidth]{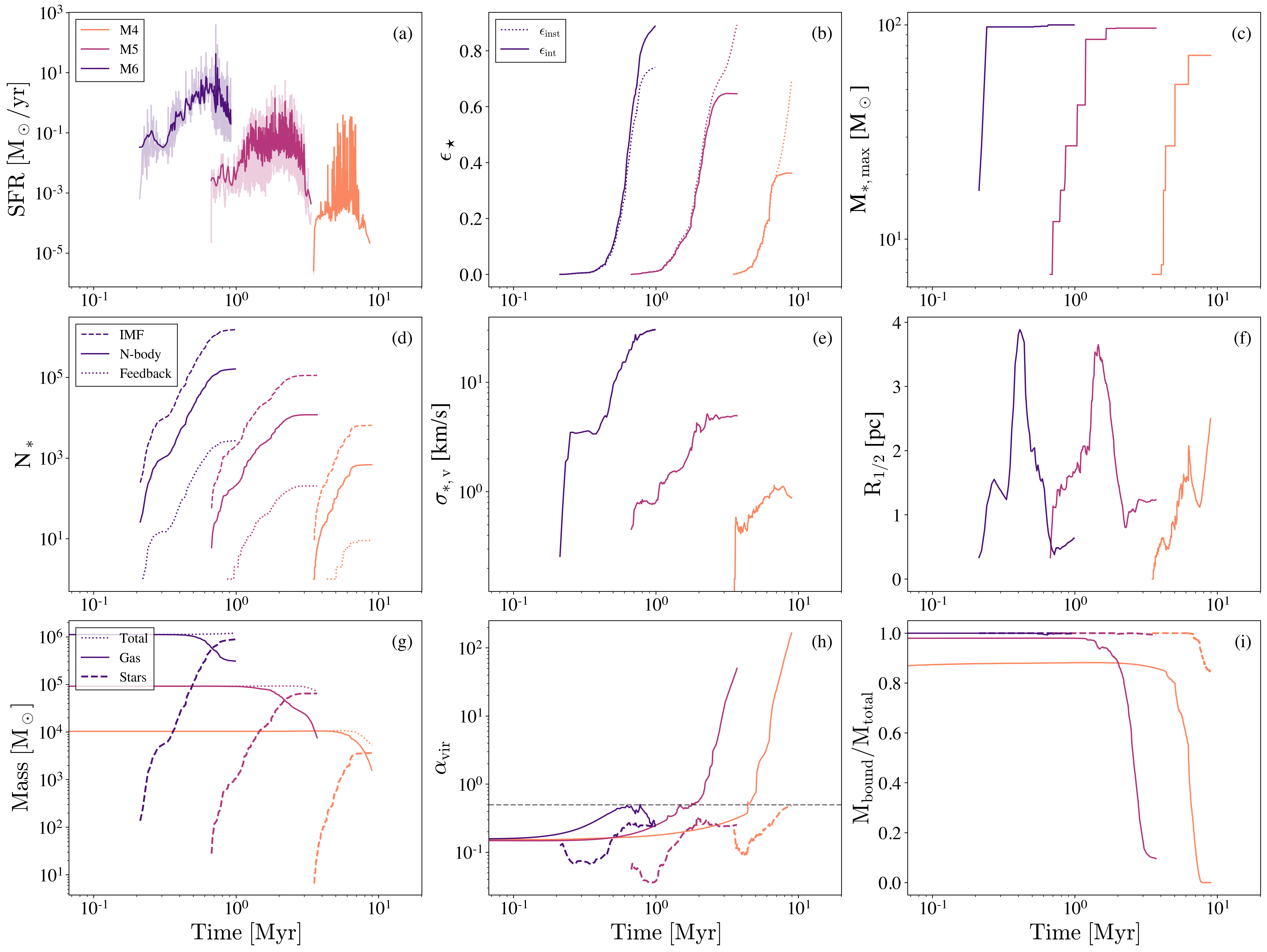}
    \caption{Global properties of the clusters and gas over time for models M4 {\em (orange)}, M5 {\em (maroon)}, and M6 {\em (blue-violet)} for comparison to Figure~\ref{fig:global_ff}, where units of free-fall time (see Table~\ref{table:test}) are used.  
    From top left to bottom right: {\em(a)} SFR, where the transparent lines show the SFR at each star formation event, and the solid lines give the SFR smoothed using a Gaussian filter with $\sigma=0.005 t_{\rm ff}$. \response{{\em(b)} Instantaneous and integrated SFEs of the clouds, where $\epsilon_{\rm inst}=M_\star/(M_{\rm gas}+M_{\rm sink}+M_\star)$ and $\epsilon_{\rm int}=M_\star/M_{\rm cloud}=\epsilon_\star$.} {\em(c)} Most massive star formed. {\em(d)} Number of formed stars. {\em Dashed line:} actual number of stars that would form from sampling the IMF given the amount of gas mass collected for star formation by sink particles. {\em Solid line:} number of stars followed in \textsc{torch} after the sampled stellar population below $4\, M_\odot$ has been agglomerated. {\em Dotted line:} number of stars above $20\, M_\odot$ on the grid that are generating feedback. The number of stars can drop due to SNe, mass loss, or exiting the grid. {\em(e)} 3D stellar velocity dispersion. {\em(f)} \response{Half-mass radius of the entire star cluster.} {\em(g)} \response{Total mass {\em (dotted line)}, mass of stars {\em (dashed line)} and gas {\em (solid line)} on the grid.} {\em(h)} Virial parameter of stars {\em (dashed line)} and gas {\em (solid line)}, where $\alpha_v=0.5$ is the equilibrium value. {\em(i)} Fraction of mass bound for stars {\em (dotted line)} and gas {\em (solid line)}.
    }
    \label{fig:global}
\end{figure*}

Figure~\ref{fig:global} reproduces Figure~\ref{fig:global_ff} using physical time rather than free-fall times to show global stellar properties over time. This demonstrates how much the duration \response{of star formation shortens while its} intensity increase\response{s} as the cloud mass increases.

\section{Stellar modifications}
\label{app:sf_mods}

\subsection{Low-mass star agglomeration}
Upwards of $10^6$ stars can be expected to form from a $10^6$~M$_{\odot}$ cloud with a peak number density of $n\approx1000\ \rm\, cm^{-3}$. Even with the best modern N-body codes, evolving this many single stars and higher order stellar systems in such a dense stellar environment with a gravity bridge from each star to the gas in a separate code is immensely computationally taxing. To reduce the strain on the N-body portion of the calculations, we chose to agglomerate all stars under a given mass into gravitational super-star particles of equivalent mass to their sum. We refer to this mass cutoff as the agglomerate mass.

\begin{figure}
\centering
    \includegraphics[width=1\columnwidth]{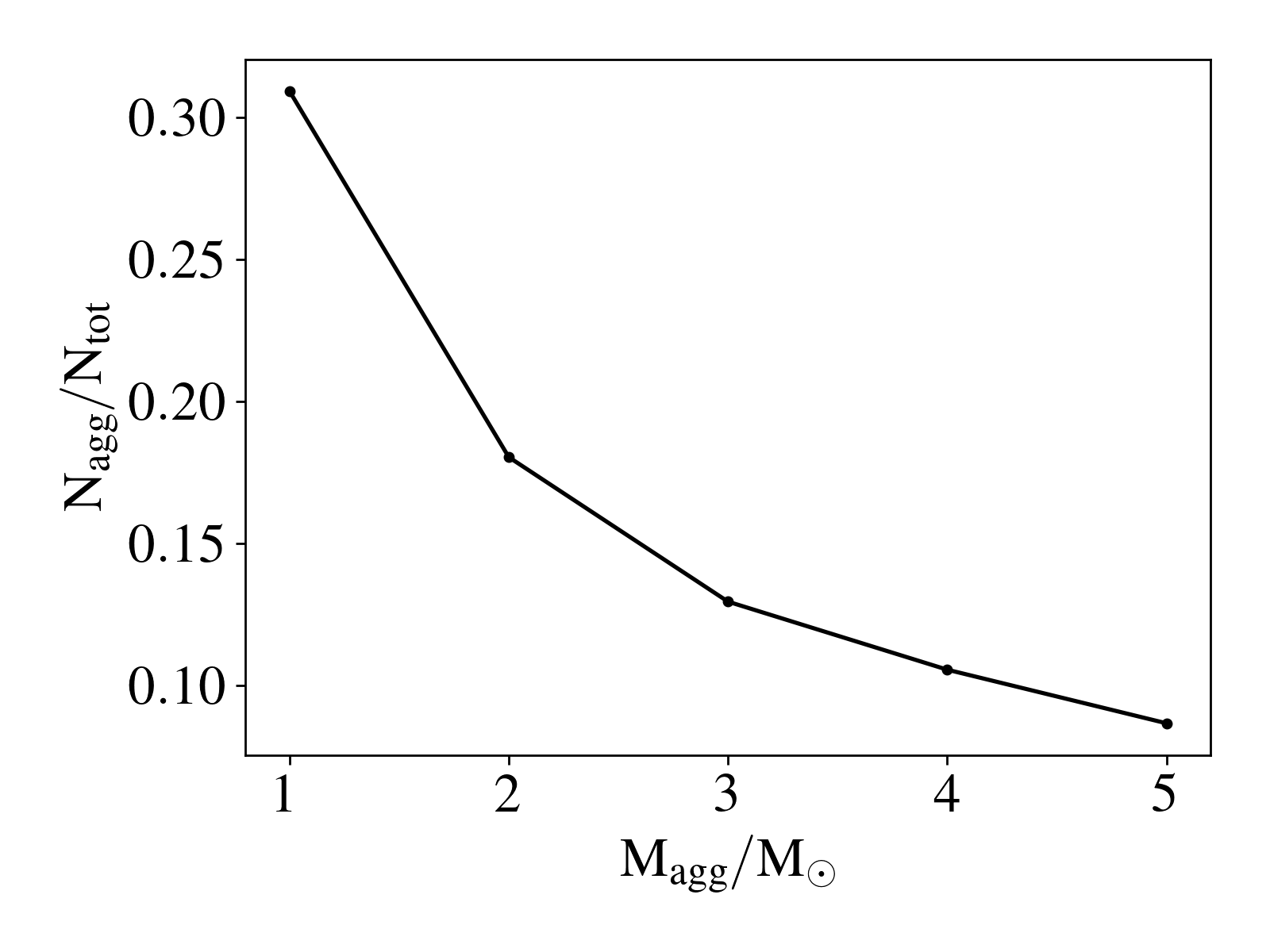}
    \caption{Fraction of the number of stars formed with agglomeration of stars below the mass on the x-axis over the total number of stars sampled by the IMF. We used $M_{\rm agg}=4\rm\, M_\odot$, which means the number of stars in the grid is $10\%$ of the number that are formed by our IMF.}
    \label{fig:m_agg}
\end{figure}

When a sink progresses through the list of stellar masses it will form, stars with masses under $M_{\rm agg}$ are put aside until the sum of their masses is above $M_{\rm agg}$. Then a star particle is formed with the summed mass. 
Figure~\ref{fig:m_agg} shows the reduction in number of stars formed in a cloud for a given agglomerate mass. For our choice of 4~M$_\odot$, we only had $~10\%$ of the stars undergoing gravitational interactions compared to the case with no agglomeration. This reduced our N-body execution time by a factor of somewhere between the $10 \log 10$ expected for the tree and $10^3$ expected for the direct N portion of the \textsc{petar} algorithm. We note that the feedback from these low-mass stars is shown in Appendix~\ref{subsec:feedback} to be negligible compared to that of the higher-mass stars, \response{and in any case the current \textsc{torch} version does not model feedback from stars $<8\rm\, M_\odot$ as we neglect jets, while the ionizing radiation from such low-mass stars is negligible. In this study, we further limited feedback to only come from stars $\ge 20\rm\, M_\odot$ as we discuss in the next subsection.} The primary missing contribution from low-mass stars physically is their mutual gravitational interactions, which could potentially lead to the ejection of some fraction of them. However, the dynamics driven by those low-mass stars is also expected to be negligible in comparison to the effect of gas and more massive stars in the cluster. \textsc{torch} simulations with no mass agglomeration were done by \citet{Cournoyer-Cloutier2023MNRAS.521.1338C}, and in analyzing the morphology of clusters they find that the gravitational effects of the gas dominate over any stellar dynamics effect for the overall evolution of the cluster while it remains embedded.

\subsection{Feedback mass limit} \label{subsec:feedback}

We limited all forms of stellar feedback---winds, radiation, and SNe---to stars above 20$\rm\, M_\odot$ instead of the value of 8$\rm\, M_\odot$ (lower bound for SN explosions) usually adopted in \textsc{torch}. This is necessary to significantly reduce the number of rays on the grid, which greatly decreases the calculation time and memory overhead for the ray-tracing algorithm. We quantify the effects of excluding radiation and winds from stars with masses below 20$\rm\, M_\odot$ by comparing the power output in the form of winds and radiation from all stars above 8$\rm\, M_\odot$ and above 20$\rm\, M_\odot$.  \response{We only allowed stars above $20\rm\, M_\odot$ to explode as SNe. Our simulations ran for $\le 10 \rm\, Myr$, which is roughly the main-sequence lifetime of a $20\rm\, M_\odot$ star. Stars under $20\rm\, M_\odot$ do not explode as SNe in the timeframe of our simulations, so excluding their SNe feedback makes no practical difference for this comparison.}

\begin{figure}
\centering
    \includegraphics[width=1\columnwidth]{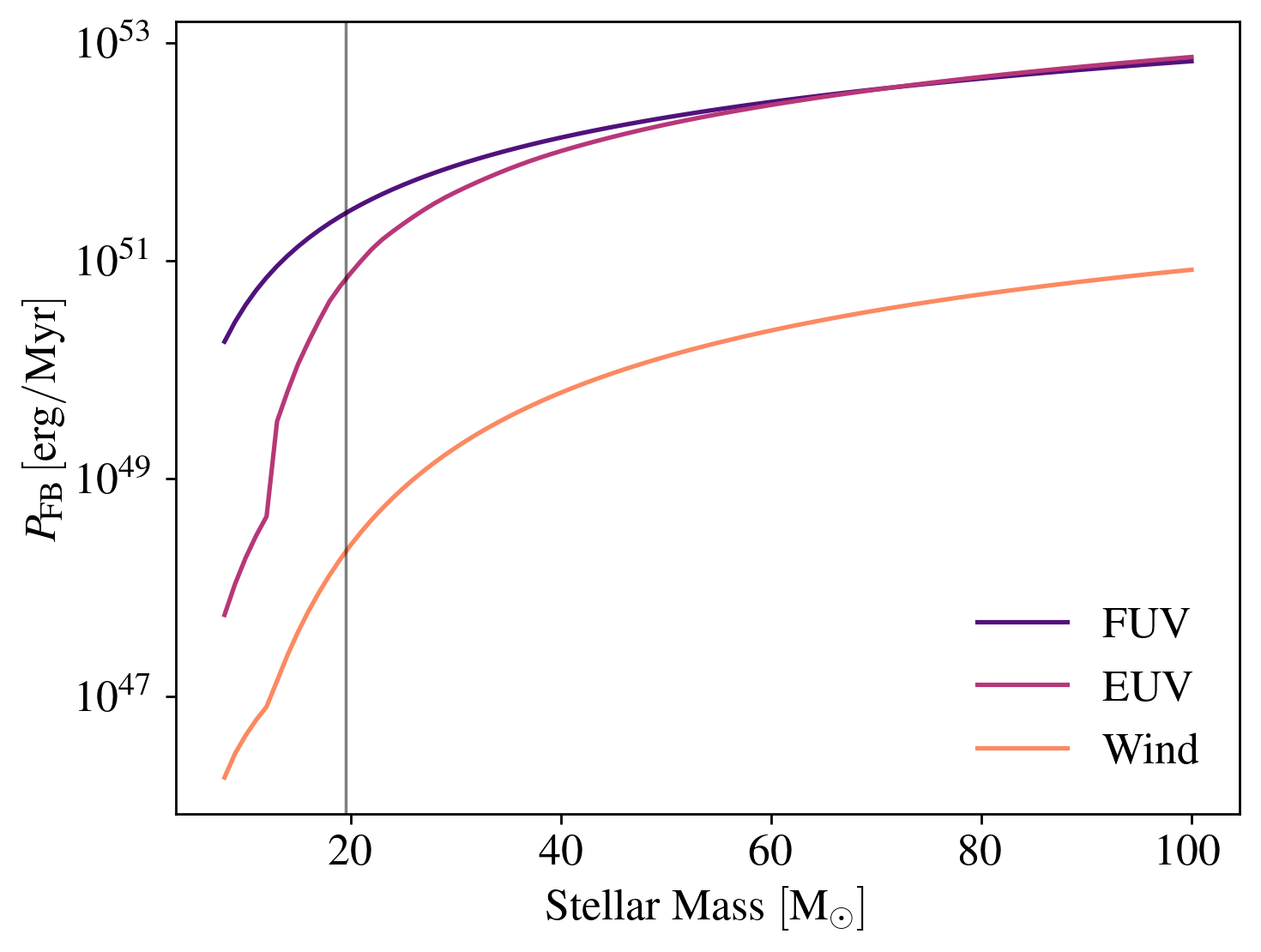}
    \caption{Power of stellar feedback in the form of winds and FUV and EUV radiation for different stellar masses. Left of the vertical line shows the amount of feedback power lost per star \response{by excluding feedback from stars below $20\, \rm\, M_\odot$.} 
    }
    \label{fig:fb_energy}
\end{figure}

The power as a function of mass in the form of EUV radiation, non-ionizing FUV radiation, and stellar wind is shown in Figure~\ref{fig:fb_energy}. We calculate these powers by taking stars from 8 to 100$\, \rm\, M_\odot$ in 1$\, \rm\, M_\odot$ increments, evolving them in \textsc{SeBa} for 1 Myr, and summing the energy output of each feedback channel. From this figure we can see that the power output of stellar winds and UV radiation is several orders of magnitude higher for stars above 20$\rm\, M_\odot$ than for stars closer to 8$\rm\, M_\odot$. Although stars in the 8--20~M$_\odot$ mass range still output a considerable amount of FUV radiation, stars above $20\, \rm M_\odot$ account for over $80\%$ of the total radiation power.

\begin{figure*}
\centering
    \includegraphics[width=0.85\textwidth]{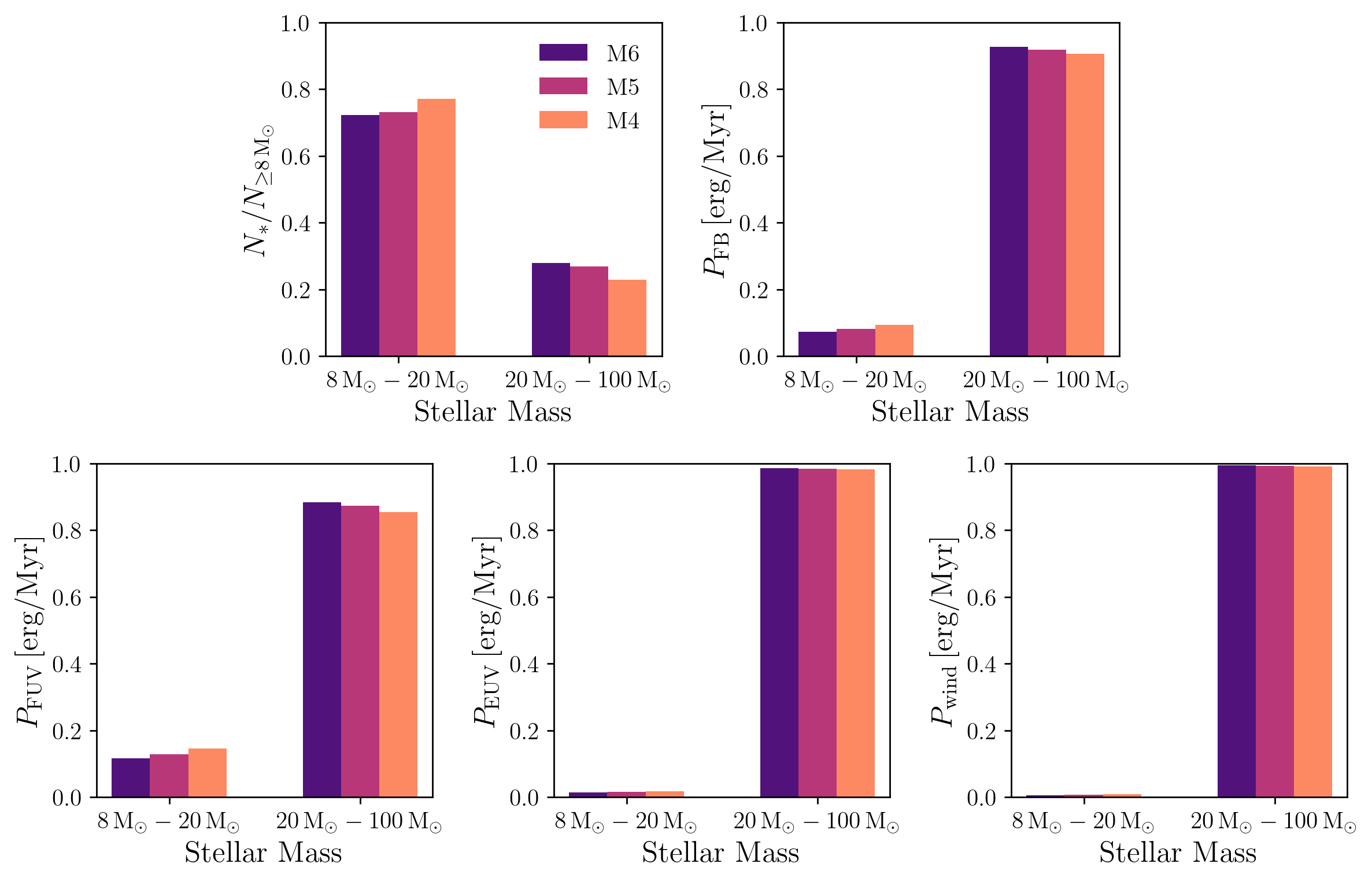}
    \caption{Power output of stellar feedback modes for each stellar mass regime. {\em (Top left):} Histogram showing the fractional stellar population of the three runs at one free-fall time, split into the mass regimes of 8--20 and 20--100~$\rm M_\odot$. {\em (Top right):} Fraction of feedback power in each mass regime.  
    {\em (Bottom):} Histograms showing the fraction of feedback power for FUV, EUV, and winds in each mass regime.
        Although there are more lower-mass stars, the feedback produced by them is less than 20\% of the total feedback energy for all stars.
    }
    \label{fig:fb_bins}
\end{figure*}

Although the feedback power is much stronger for stars above 20$\rm\, M_\odot$, stars with masses 8--20$\rm\, M_\odot$ greatly outnumber them. To find the ratio of feedback power for stars below and above 20~$\rm\, M_\odot$, we convolve the number of stars of each mass with the power output for each stellar mass (Fig.~\ref{fig:fb_bins}).  In the top left histogram, we show the ratio of stars with mass 8--20~$\rm\, M_\odot$ to stars with mass 20--100~$\rm\, M_\odot$ in all three simulations, sampled at their respective initial free-fall times. All three runs have more stars in the lower-mass bin. In the top right plot, we show the ratio of total stellar feedback power $P_{\rm FB}$ (excluding SNe) for the stars in the two mass bins considered. We can see that although the lower-mass stars outnumber the higher-mass stars, the higher-mass stars still account for $>80\%$ of the total stellar feedback energy. This shows that only including feedback from stars above 20~$\rm\, M_\odot$ still retains almost all of the feedback energy produced after the formation of all three star clusters.

The bottom panel in Figure~\ref{fig:fb_bins} shows the feedback power per mass bin for each separate feedback process. For the EUV radiation and wind feedback, the low-mass stars contribute practically nothing to the feedback energy in comparison to the high-mass stars. The FUV feedback of low-mass stars is not negligible, but is still well below $20\%$ of the total FUV feedback energy from all stars.

\subsection{Mass-loading stellar winds}
In \textsc{torch}, the stellar wind feedback implementation is inspired by \citet{Simpson2015ApJ...809...69S}, using a method of momentum injection, the details of which can be found in \citet{2020Wall}. The energy of the cells within the wind injection radius of the star is increased based on the mechanical luminosity of the wind $L_{\rm w}=(1/2)\Dot{M}v_{\rm w}^2$, where $\Dot{M}$ is the stellar mass loss rate \citep{Vink2000A&A...362..295V} and $v_{\rm w}$ is the terminal wind velocity \citep{Kudritzki2000ARA&A..38..613K}. The wind injection radius is set by comparing the cell width $\Delta x$ to the wind termination shock radius \citep{Weaver1977ApJ...218..377W}
\begin{equation}
    R_{\rm w} = 0.74 \left(\frac{\Dot{M}}{\rho_0}\right)^{3/10} v_{\rm w}^{1/10} t_{\rm w}^{2/5},
\end{equation}
where $\rho_0$ is the background density and $t_{\rm w}$ is the \response{age of the wind-blowing star at the given time step.} If $R_{\rm w}<\Delta x$ the injection radius is set to $\Delta x$, otherwise it is set to a maximum value of $6\sqrt{3}\Delta x$, at which we have found that spherical winds are well resolved. Momentum and energy are conserved when injecting stellar winds.

Within a stellar wind bubble, in dense clumpy regions of star formation such as the ones in our simulations, material will be swept up into the flow of the hot bubble by mass loading processes such as photoevaporation and hydrodynamic ablation \citep{Dyson1992ApL&C..28..301D,Hartquist1996Ap&SS.245..263H,Pittard2001A&A...373.1043P,Lancaster2021ApJ...914...89L}. With enough mass loading, the density increase will result in much more efficient cooling and create momentum-driven rather than energy-driven bubbles. The amount of mass-loading in the case of hydrodynamic ablation depends on the prevalence of dense clumps within the wind region as well as the Mach number $M$ of the flow around the clump. With a supersonic flow, the mass-loading rate saturates. With a subsonic flow, the mass-loading rate is proportional to $M^{4/3}$ \citep{Smith1984MNRAS.211..679S,Hartquist1986MNRAS.221..715H}. Accounting for mass loading in stellar wind models has been shown to successfully reproduce the kinematic properties of the observed stellar wind bubble of the Wolf-Rayet star RCW 58 \citep{Arthur1993MNRAS.261..425A,Arthur1996A&A...313..897A,Arthur2007ASSP....1..183A}.

Simply injecting winds at $v_{\rm w}$ does not account for these mass-loading processes and results in unphysically hot bubbles. Therefore, we chose a lower temperature target for our bubbles and lowered the wind velocity $v_{\rm w}$ such that the final temperature of the wind bubble is the correct one. We conserved momentum and energy when injecting stellar winds, so while lowering the wind velocity, we also infused correspondingly more mass into the bubble than the stellar mass loss calculated. This mass is not taken off the grid elsewhere, meaning mass was not entirely conserved. \response{At $1.25\,t_{\rm ff}$, the total amount of mass that has been injected due to mass-loading as a fraction of initial cloud mass is 0.04, 0.02, and 0.01 for M4, M5, and M6.}

Observed circumstellar bubbles cooled by suspected mass loading have been seen with temperatures as low as $T_{\rm b}\approx 1.1\times 10^6\rm\, K$ in the S308 bubble \citep{Chu2003ApJ...599.1189C}. The spectra of the NGC 6888 bubble indicates a dominant component almost as cool, with $T_{\rm b}=1.5\times 10^6\rm\, K$ \citep{Bochkarev1988Natur.332..518B,Wrigge1994A&A...286..219W,Wrigge1999A&A...343..599W}.

In the simulations presented here, we heavily mass loaded the stellar winds to achieve a lower than observed bubble temperature of $T_{\rm b}=3\times 10^5$~K. 
This temperature is at the peak of the cooling curve, so the shocked wind rapidly cools, resulting in smaller, cooler, momentum-driven bubbles instead of hot bubbles filled with $10^6$~K gas. We chose to do so because the high sound speeds in hot wind bubbles lower the Courant time step significantly, making the computation impractical. Since we do not follow X-rays through ray-tracing, having cooler bubbles is adequate. Bubbles at this temperature also do not affect the ionization of the surrounding gas. The primary action of wind feedback during cluster formation is to clear out dense regions of gas so that radiatively ionized H~{\sc ii} regions can expand. The only hot gas ($\ge10^6$ K) on the grid comes from SNe. Capping the temperature of gas on the grid at $3 \times 10^5$~K until SNe occur significantly speeds up the simulations.

\begin{figure}
\centering
    \includegraphics[width=1\columnwidth]{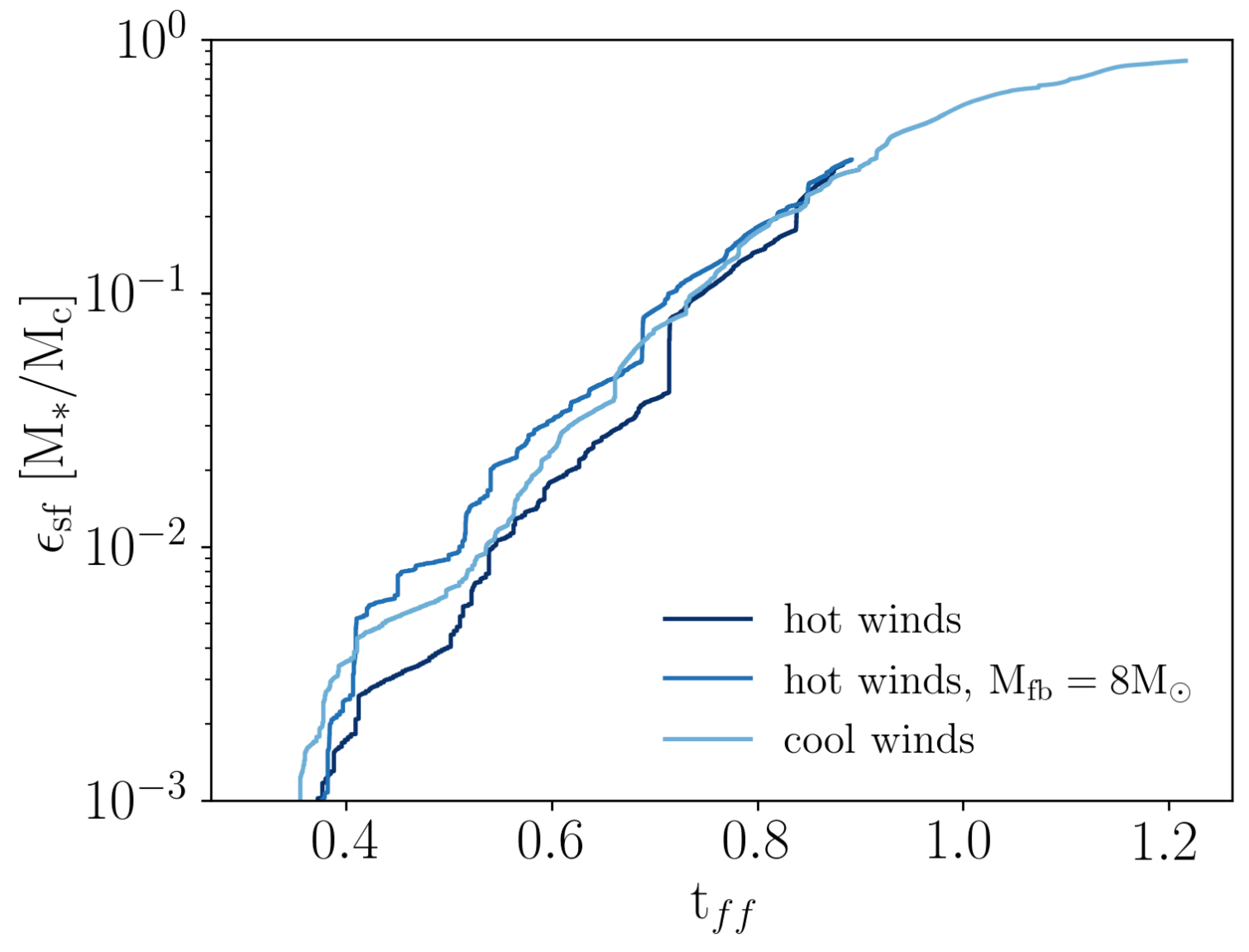}
    \caption{Star formation efficiency over time for the fiducial M6 cloud with $T_{\rm w}=300,000\rm\, K$ and $M_{\rm feedback}\ge 20\rm\, M_\odot$ (\textit{cool winds}), the M6 cloud with $T_{\rm w}=5,000,000\rm\, K$ and $M_{\rm feedback}\ge 20\rm\, M_\odot$ (\textit{hot winds}), and the M6 cloud with $T_{\rm w}=5,000,000\rm\, K$ and $M_{\rm feedback}\ge 8\rm\, M_\odot$ (\textit{hot winds, $M_{\rm fb}=8\rm\, M_\odot$}).}
    \label{fig:sfe_m6}
\end{figure}

\subsection{Effect on star formation efficiency}

Limiting the temperature of stellar winds and only modelling feedback for stars above $20\rm\, M_\odot$ could potentially lead to un-physical runaway star formation. To test this, we re-ran the M6 model at early times to see if these two approximations are the cause for the extremely high SFE of $85\%$. For the first new M6 run we raised the wind temperature from $3 \times 10^5\rm\, K$ to $5 \times 10^6\rm\, K$. For the second test, we both raised the wind temperature and modelled feedback for all stars above $8\rm\, M_\odot$. The SFE over time for the fiducial M6 run with our standard approximations and the new M6 models are shown in Figure~\ref{fig:sfe_m6}.

The two runs without the approximations that reduce the strength of the stellar feedback have similar SFEs as the M6 model with the aforementioned approximations. This validates our approximations and supports our argument that the high SFE in model M6 is not an artifact of under-estimating the strength of stellar feedback.

\end{appendix}

\end{document}